\documentclass[12pt]{article}
\usepackage{psfrag,epsfig,graphicx,graphics}

\usepackage{amssymb}

\newcommand{\widl}{0.7\columnwidth}
\newcommand{\wid}{0.6\columnwidth}
\newcommand{\widm}{0.5\columnwidth}
\newcommand{\wids}{0.45\columnwidth}
\newcommand{\widss}{0.37\columnwidth}

\newcommand{\scalingb}{3.2cm}
\newcommand{\scalingl}{4cm}

\def\slashchar#1{\setbox0=\hbox{$#1$}
   \dimen0=\wd0
   \setbox1=\hbox{/} \dimen1=\wd1
   \ifdim\dimen0>\dimen1
      \rlap{\hbox to \dimen0{\hfil/\hfil}}
      #1
   \else
      \rlap{\hbox to \dimen1{\hfil$#1$\hfil}}
      /
   \fi}

\def\bea{\begin{eqnarray}}
\def\beqa{\begin{eqnarray}}
\def\eea{\end{eqnarray}}
\def\eqa{\end{eqnarray}}
\def\beas{\begin{eqnarray*}}
\def\eeas{\end{eqnarray*}}
\def\beqas{\begin{eqnarray*}}
\def\eqas{\end{eqnarray*}}
\def\beq{\begin{equation}}
\def\eeq{\end{equation}}
\def\beqd{\begin{displaymath}}
\def\eeqd{\end{displaymath}}
\def\eqd{\end{displaymath}}

\def\beeq{\begin{eqnarray}} \def\eeeq{\end{eqnarray}}

\newcommand{\la}{\lambda}

\newcommand{\be}{\begin{equation}}
\newcommand{\ee}{\end{equation}}
\newcommand{\eq}{\end{equation}}
\newcommand{\rb}{\underline{r}}
\newcommand{\kb}{\underline{k}}

\newcommand{\pb}{\underline{p}}

\newcommand{\Qb}{\underline{Q}}
\newcommand{\epsb}{\underline{\epsilon}}
\newcommand{\zb}{\bar{z}}

\newcommand{\aab}{\bar{a}}
\newcommand{\lr}{\leftrightarrow}
\newcommand{\eps}{\epsilon}
\newcommand{\intfeyn}{\int\limits_0^1}
\newcommand{\as}{\bar\alpha_s}

\begin{document}

\begin{flushright}
LPT-07-11\\  CPHT-RR002.0107\\
  hep-ph/0703166
\end{flushright}

\vspace{\baselineskip}

\begin{center}
\textbf{\LARGE Diffractive production of two $\rho^0_L$ mesons in $e^+e^-$ collisions
}\\

\vspace{3\baselineskip}
\vspace{3\baselineskip}

\vspace{3\baselineskip}
{\large
M. Segond,$^a$
 L. Szymanowski$^{a,b,c,d}$ and S. Wallon$^a$
}
\\

\vspace{1\baselineskip}
${}^a$\,LPT, Universit\'e Paris-Sud -CNRS, 91405-Orsay, France  \\[0.5\baselineskip]
${}^b$\,Soltan Institute for Nuclear Studies, Warsaw, Poland
\\[0.5\baselineskip]
${}^c$\,Universit\'e  de Li\`ege,  B4000  Li\`ege, Belgium\\[0.5\baselineskip]
${}^d$\,CPHT,
\'Ecole Polytechnique-CNRS, 91128 Palaiseau, France \\[0.5\baselineskip]

\vspace{5\baselineskip}
\textbf{Abstract}\\
\vspace{1\baselineskip}
\parbox{0.9\textwidth}{We present an estimate
of the cross-section for the exclusive production of a $\rho_L^0$-meson pair in  $e^+e^-$ scattering,
which will be studied in the future high energy International Linear Collider. For this aim, we 
complete calculations of the Born order approximation of the amplitudes $\gamma^{*}_{L,T} (Q_1^2)\gamma^{ *}_{L,T}(Q_2^2) \to \rho_L^0 \rho_L^0,$ for arbitrary
polarization of virtual photons and longitudinally polarized mesons, in the kinematical region $s \gg -t,
\,Q_1^2 \,, Q_2^2.$
These processes are completely calculable in the hard region $Q_1^2 \,, Q_2^2 \gg
\Lambda^2_{QCD}$ and we perform most of the calculations in an analytical way.
The resulting cross-section turns out to be large enough for this process to
be measurable with foreseen luminosity and energy, for $Q_1^2$ and $Q_2^2$ in the range of a
few ${\rm GeV^2}.$
}

\end{center}

\eject

\medskip
\noindent
\section{ Introduction}
\setcounter{equation}{0}
\setcounter{figure}{0}

The next generation of $e^+e^--$colliders will offer a possibility of
clean testing of QCD dynamics.
 By selecting events in which two vector mesons are produced
with large rapidity gap, through scattering of two highly virtual photons,
one is getting access to the kinematical regime in which the perturbative
approach is justified. If additionally one selects the events with
comparable photon virtualities, the perturbative Regge dynamics of QCD
 of the BFKL \cite{bfkl} type should dominate with respect to the
conventional partonic evolution of DGLAP \cite{dglap} type.
Several studies of BFKL dynamics have been performed at the level of the total cross-section
\cite{bfklinc,royon}. 

In the paper \cite{2jpsi} the diffractive production of two $J/\Psi$ mesons was studied as
a promising
probe
of the BFKL effects. 
 Recently, we have advocated \cite{conf, psw, epsw} that  the  electroproduction
of two  $\rho-$mesons in the $\gamma^* \gamma^*$ offers the same advantages.
 In this case the virtualities of the scattered photons play the role of the hard
scales.
A first step in this direction was made by considering this
process with  longitudinally polarized photons and
$\rho-$mesons,
 \beq
\label{processLL}
\gamma^*_L(q_1)\;\gamma^*_L(q_2) \to \rho_L^0(k_1)  \;\rho_L^0(k_2)\,,
\ee
for arbitrary values of $t=(q_1-k_1)^2,$ with $s \gg -t.$
 The choice of longitudinal polarizations of both the
scattered photons and produced vector mesons was dictated by the
fact that this configuration of the lowest twist-2 gives the dominant
contribution in the powers of the hard scale $Q^2,$ when $Q_1^2 \sim Q_2^2
\sim Q^2$.

\begin{figure}[htbp]
\begin{center}
\epsfig{file=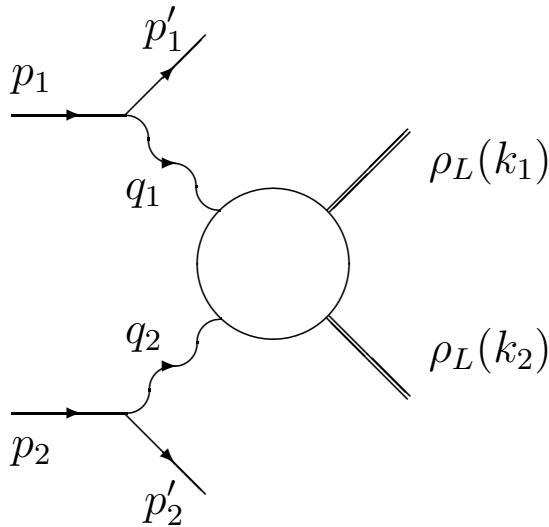,width=\wids}
\end{center}
\caption{\small Amplitude for the process $e^+ \, e^- \to e^+ \, e^-  \, \rho^0_L \, \rho^0_L$.}
\label{Figeeprocess}
\end{figure}

The aim of this study is to complete the Born order evaluation of the cross-section of the process
\beq
\label{eeprocess}
e^+ e^- \to e^+ e^- \rho_L^0 \rho_L^0\,,
\eq
illustrated in Fig.\ref{Figeeprocess}.
We calculated contributions  
with remaining combinations of polarizations of virtual photons necessary
to obtain all helicity amplitudes of the processes 
\beq
\label{process}
\gamma^*_{L,T}(q_1)\;\gamma^*_{L,T}(q_2) \to \rho^0_L(k_1)  \;\rho^0_L(k_2)\,.
\ee
Let us note, that the double tagging of final leptons gives in particular the possibility to
separate the contributions of various photon polarizations,
entering in (\ref{eeprocess}) and thus to study
the corresponding parts of the cross-sections  which are computed in this paper.
 We are focusing here on the high-energy limit in which $t-$channel gluonic
exchanges dominate. 
On the other hand, in the description of the process (\ref{eeprocess}), there is a potential possibility
that also contributions with rather small $s_{\gamma^* \gamma^*}$
have to be taken into account. In this case one should 
include in principle both quark and gluon exchanges.
The contribution of quark exchange was analyzed in \cite{gdatda}. This quark-box contribution
is investigated in subsection \ref{subdominant}.

We will also not consider here the case of transversally
polarized $\rho-$mesons. It would require to deal with
possible breaking of QCD factorization \cite{break}, although
 a method to overcome this problem has been proposed
 \cite{ryskin}.

 The BFKL enhancement was studied
for $t=0$ in \cite{epsw} and \cite{IvanovPapa}. In this latter case, the
peculiar value $t=0$ automatically selects the longitudinally polarized
photon.
A dedicate study for arbitrary value of $t$ should thus be performed to get an evaluation of BFKL enhancement
effects of the Born order evaluation performed in the present
paper for transversally polarized photon. This problem will not be addressed here.

\section{Kinematics}
\label{kinematics}
\setcounter{equation}{0}

\begin{figure}[htb]
\psfrag{r1}[cc][cc]{$\quad\rho(k_1)$}
\psfrag{r2}[cc][cc]{$\quad\rho(k_2)$}
\psfrag{p1}[cc][cc]{$\slashchar{q}'_1$}
\psfrag{p2}[cc][cc]{$\slashchar{q}'_2$}
\psfrag{q1}[cc][cc]{$q_1$}
\psfrag{q2}[cc][cc]{$q_2$}
\psfrag{Da}[cc][cc]{DA}
\psfrag{HDA}[cc][cc]{$M_H$}
\psfrag{M}[cc][cc]{$M$}
\centerline{\scalebox{1}
{
$\begin{array}{cccc}
\raisebox{-0.44 \totalheight}{\epsfig{file=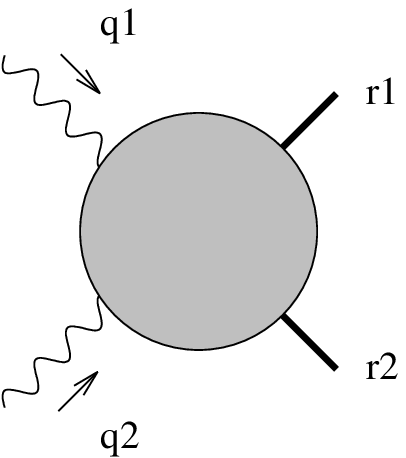,width=\scalingl}}&=&
\raisebox{-0.44\totalheight}{\epsfig{file=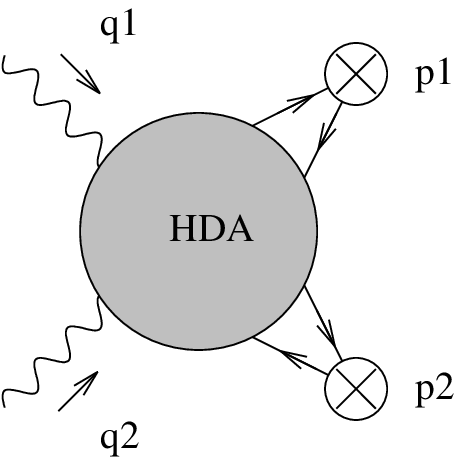,width=\scalingl}}
& \begin{array}{c}
\raisebox{0.4 \totalheight}
{\psfrag{r}[cc][cc]{$\quad\rho(k_1)$}
\psfrag{pf}[cc][cc]{$\slashchar{q}'_2$}
\epsfig{file=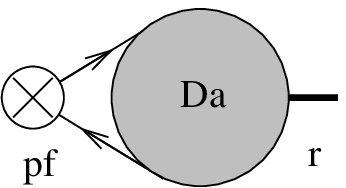,width=\scalingl}}\\
\raisebox{0.1 \totalheight}
{\psfrag{r}[cc][cc]{$\quad\rho(k_2)$}
\psfrag{pf}[cc][cc]{$\slashchar{q}'_1$}
\epsfig{file=DA.eps,width=\scalingl}}
\end{array}
\end{array}
$}}
\caption{\small The amplitude of the process $\gamma^*(Q_1) \gamma^*(Q_2) \to \rho^0_L (k_{1})\rho^0_L(k_{2})$ with the collinear factorization in the $q \bar{q} \rho$ vertex. \label{FigCollinear}}
\end{figure}

\begin{figure}[htbp]
\begin{center}
\epsfig{file=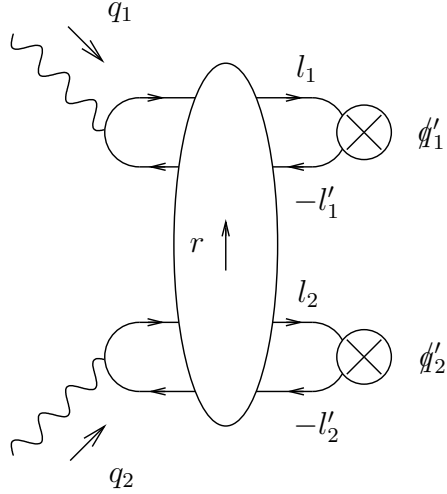,width=\widss}
\end{center}
\caption{\small The amplitude $M_H$
in the impact representation. The vertical blob symbolizes the interaction of two $q \bar{q}$ dipoles
through gluon exchanges at high $s.$}
\label{Figprocess}
\end{figure}

The measurable cross section for the process (\ref{eeprocess})
of Fig.\ref{Figeeprocess}
is related to the amplitude of the process (\ref{process}), illustrated in Fig.\ref{FigCollinear},
 through the usual
  flux factors for respectively transversally and longitudinally
polarized photons
\beq
t(y_i) = \frac{1 + (1-y_i)^2}{2}, \quad l(y_i)=1-y_i  \,,
\eq
where $y_{i}\, (i=1,2)$ are the longitudinal momentum fractions of the bremsstrahlung photons with
respect to the incoming leptons.
This relation reads \cite{budnev}
\bea
\label{eesigma}
&& \frac{d\sigma(e^+ e^- \to e^+ e^- \rho_L^0 \rho_L^0)}{dy_{1}dy_{2}dQ_{1}^2dQ_{2}^2} \\
&& = \frac{1}{y_1y_2\,Q_1^2Q_{2}^2}\left(\frac{\alpha}{\pi}\right)^2
\left[ l(y_1)\, l(y_2)\, \sigma(\gamma^*_L\gamma^*_L\to \rho_L^0 \rho_L^0) + \,  t(y_1)\, l(y_2)\, \sigma(\gamma^*_T\gamma^*_L\to \rho_L^0 \rho_L^0)\right. \nonumber \\
&&+ \, \left.l(y_1)\, t(y_2)\, \sigma(\gamma^*_L\gamma^*_T\to \rho_L^0 \rho_L^0) +t(y_1)\, t(y_2)\, \sigma(\gamma^*_T\gamma^*_T\to \rho_L^0 \rho_L^0)\right]\;. \nonumber
\eea
The presence of hard scales $Q_i^2$ permits us to apply the collinear 
approximation at each $q \bar{q} \rho-$meson vertex, and the use of distribution amplitude (DA) for describing the $q \bar{q}$
content
of the $\rho$ mesons, as illustrated in Fig.\ref{FigCollinear}.
In this paper, except for section \ref{subdominant}, the amplitude $M_H$ will be described using the impact representation, valid at high energy, as illustrated  in Fig.\ref{Figprocess}.
\psfrag{p1}[cc][cc]{$\slashchar{q}'_1$}
\psfrag{p2}[cc][cc]{$\slashchar{q}'_2$}
\psfrag{q1}[cc][cc]{$q_1$}
\psfrag{q2}[cc][cc]{$q_2$}
\psfrag{l1}[cc][cc]{$l_1$}
\psfrag{l1p}[cc][cc]{$-l_1'$}
\psfrag{l2}[cc][cc]{$l_2$}
\psfrag{l2p}[cc][cc]{$-l_2'$}
\psfrag{r}[cc][cc]{$r$}

Let us introduce
 two light-like
Sudakov vectors $q'_1$ and $q'_2$ which form a natural basis
for two scattered virtual photons, which
satisfy
$2 q'_1 \cdot q'_2 \equiv s \sim 2 q_1 \cdot q_2.$ The usual
$s_{\gamma^*\gamma^*}$ is related to the auxiliary useful variable $s$ by $s_{\gamma^*\gamma^*}=s
-Q_1^2-Q_2^2.$ The momentum transfer in the $t-$channel is $r= k_1 -q_1.$
In this
basis, the
incoming photon momenta read
\beq
\label{inmom}
q_1= q'_1 - \frac{Q_1^2}{s} q_2' \quad {\rm and} \quad
q_2 =q'_2 - \frac{Q_2^2}{s} q_1'\,.
\eq
The polarization vectors of longitudinally polarized photons are
\beq
\label{inpolL}
\epsilon^{L(1)}_{\mu}= \frac{q_{1 \mu}}{Q_1} + \frac{2 Q_1}{s} q'_{2\mu} \quad {\rm and} \quad
\epsilon^{L(2)}_{\mu}= \frac{q_{2 \mu}}{Q_2} + \frac{2 Q_2}{s} q'_{1\mu}\;,
\eq
with $\epsilon_{L(i)}^2=1$ and
$q_i \cdot \epsilon_{L(i)}=0$, whereas 
the polarization vectors of transversally polarized photons are
  two dimensional transverse vectors
satisfying  $\epsilon_{T(i)}^2=-1 \,\, (i=1,2)$ and
$q_i \cdot \epsilon_{T(i)}=0.$

We label the momentum of the quarks and antiquarks entering the meson
wave functions as $l_1$ and $l'_1$ for the upper part of the diagram
and $l_2$ and $l'_2$ for the lower part (see Fig.\ref{Figprocess}).

In the basis (\ref{inmom}), the vector meson momenta can be expanded in
the form
\bea
\label{rotk1k2}
&& k_1 = \alpha (k_1) \, q_1' + \frac{\rb^2}{\alpha (k_1) \, s} q_2' + r_\perp \;,
\nonumber  \\
&& k_2 = \beta  (k_2)\, q_2' + \frac{\rb^2}{\beta (k_2) \, s} q_1' - r_\perp \,.
\eea
Note that our convention is such that for any tranverse vector
$v_\perp$ in Minkowski space, $\underline{v}$ denotes its euclidean
form. In the following, we will treat the $\rho$ meson as being massless. $\alpha$ and $\beta$
are very close to  unity (explicit expressions can be found in \cite{psw}), and reads
\bea
\label{defalphabeta}
&&\alpha (k_1) \simeq 1 -\frac{Q_2^2 + \rb^2}{s} + O\left(\frac{1}{s^2}\right)\,,
\nonumber \\
&& \beta (k_2) \simeq 1 -\frac{Q_1^2 + \rb^2}{s} + O\left(\frac{1}{s^2}\right)\,,
\eea
where $\rb^2 = -r_\perp^2$. They will be replaced by 1 in the phenomenological applications of sections \ref{cross}
and \ref{crossee}.
In this decomposition, it is straightforward to relate $t=r^2$ to $\rb^2$  through
the approximate relation 
\beq
\label{tfonctionr}
t \sim -\frac{Q_1^2 \, Q_2^2}{s} -\rb^2 \left(1+ \frac{Q_1^2}{s}+\frac{Q_2^2}{s} +\frac{\rb^2}{s} \right)\,
\eq
 (see \cite{psw} for an exact relation).
From Eq.(\ref{tfonctionr}) the threshold for $|t|$ is given by
$|t|_{min}=Q_1^2 Q_2^2/s\,,$ corresponding to $r_\perp = 0.$
In the kinematical range we are interested in, the relation
(\ref{tfonctionr})
can be approximated as  $\rb^2=-t,$ as usually in the Regge limit.

  The links with the $e^+ e^-$ process can be made by using the
  same Sudakov basis for the two incoming leptons:
\beq
\label{sudakovlepton}
p_1 = \frac{1}{y_1} q_1' + y_1 \frac{\pb_1^2}{s} q_2' + p_{\perp 1} \quad {\rm and} \quad p_2 = \frac{1}{y_2} q_2' + y_2 \frac{\pb_2^2}{s} q_1' + p_{\perp 2}\,,\quad {\rm with} \quad
\pb_i^2= \frac{1-y_i}{y_i^2} Q_i^2\,.
\eq
Thus, one gets
$$s_{e^+e^-}=\frac{s}{y_1 y_2} \left(1 + \frac{(1-y_1)(1-y_2) Q_1^2 Q_2^2}{s^2}\right) -2 \pb_1 \cdot \pb_2 \,.$$
In the rest of the paper, since we keep only the dominant $s$ contribution, we  use the approximate relation  $s_{e^+e^-} \sim s/(y_1 y_2).$

\section{Impact representation}
\label{impact}

\setcounter{equation}{0}

\begin{figure}[htbp]
\begin{center}
\epsfig{file=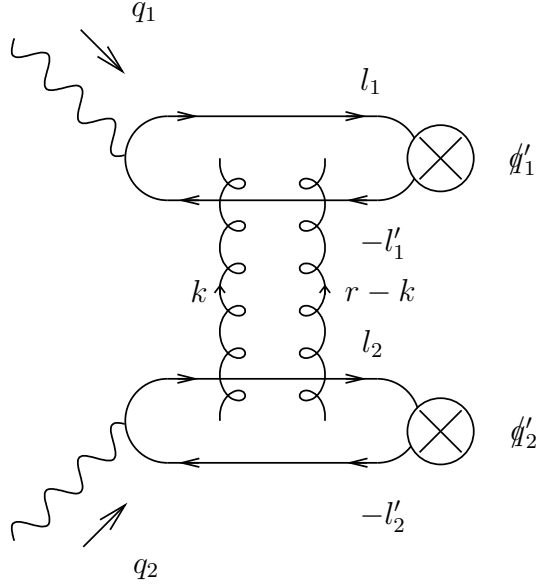,width=\wids}
\end{center}
\caption{\small Amplitude $M_H$
at Born order. The $t$-channel gluons are attached to the quark lines in all
possible ways.}
\label{Figborn}
\end{figure}

The impact factor representation of the scattering amplitude for the process
(\ref{process}) has the form (see Fig.\ref{Figborn})
\psfrag{k}[cc][cc]{$k$}
\psfrag{rmk}[cc][cc]{$r-k$}
\be
\label{M}
{\cal M} = is\;\int\;\frac{d^2\,\kb}{(2\pi)^4\kb^2\,(\rb -\kb)^2}
{\cal J}^{\gamma^*_{L,T}(q_1) \to \rho^0_L(k_1)}(\kb,\rb -\kb)\;
{\cal J}^{\gamma^*_{L,T}(q_2) \to \rho^0_L(k_2)}(-\kb,-\rb +\kb)\,,
\ee
where ${\cal J}^{\gamma^*_{L,T}(q_1) \to \rho^0_L(k_1)}(\kb,\rb -\kb)$
(${\cal J}^{\gamma^*_{L,T}(q_2) \to \rho^0_L(k_2)}(\kb,\rb -\kb)$)
are the impact factors corresponding  to the
transition of
$\gamma^*_{L,T}(q_1)\to \rho^0_L(k_1)$ ($\gamma^*_{L,T}(q_2)\to \rho^0_L(k_2)$) via the
$t-$channel exchange of two gluons.
 The amplitude (\ref{M}) calculated in Born order depends linearly on $s$ (or $s_{\gamma^*\gamma^*}$ when neglecting terms of order $Q_i^2/s$) as the impact factors are $s$-independent.

Calculations of the impact factors in the Born approximation\footnote{Now, the forward impact factor of $\gamma^*_L(Q^2) \to \rho^0_L$
transition has been calculated at the next-to-leading order accuracy in
\cite{ivanov}.}are standard
\cite{ginzburg}. They are obtained by assuming the collinear approximation 
at each $q \bar{q} \rho-$meson vertex. Projecting the (anti)quark momenta on the Sudakov basis $q_1',\, q_2',$
\bea
\label{l}
&& l_1 = z_1 q'_1 + l_{\perp 1} + z_1 r_\perp - \frac{(l_{\perp 1} + z_1
r_\perp)^2}{z_1
s} q'_2 \;,\nonumber \\
&& l'_1 = \zb_1 q'_1 - l_{\perp 1} + \zb_1 r_\perp - \frac{(-l_{\perp 1} +
\zb_1
r_\perp)^2}{\zb_1 s} q'_2 \;, \nonumber \\
&& l_2 = z_2 q'_2 + l_{\perp 2} - z_2 r_\perp - \frac{(l_{\perp 2} - z_2
r_\perp)^2}{z_2
s} q'_1 \;, \nonumber \\
&& l'_2 = \zb_2 q'_2 - l_{\perp 2} - \zb_2 r_\perp - \frac{(-l_{\perp 2} -
\zb_2 r_\perp)^2}{\zb_2 s} q'_1 \,,
\eea
we put the relative momentum $l_{i\perp}$  to zero.
For longitudinally polarized photons the impact factor reads 
\beq
\label{ifL}
\hspace{-.8cm}{\cal J}^{\gamma^*_L(q_i) \to \rho_L(k_i)}(\kb,\rb -\kb)
= 8 \pi^2 \alpha_s
\frac{e}{\sqrt{2}} \frac{\delta^{a b}}{2 N_c} Q_i \, f_\rho \, \alpha(k_i) \int\limits_0^1 dz_i z_i \, \zb_i \, \phi(z_i)
\rm{P_P(z_i,\kb,\rb,\mu_i)} \, ,
\eq
where the expression
\beq
\label{defPP}
\rm{P_P(z_i,\kb,\rb,\mu_i)}=
 \frac{1}{z_i^2\rb^2 + \mu_i^2} +
\frac{1}{\zb_i^2\rb^2 + \mu_i^2}
 - \frac{1}{(z_i\rb -\kb)^2 + \mu_i^2} - \frac{1}{(\zb_i\rb
-\kb)^2
+ \mu_i^2}
\eq
originates from the impact factor of quark pair production from a
longitudinally polarized photon.

For transversally polarized photons, one obtains
\bea
\label{ifT}
&&\hspace{-.8cm}{\cal J}^{\gamma^*_T(q_i) \to \rho_L(k_i)}(\kb,\rb -\kb)
\nonumber \\
&&\hspace{-.8cm}= 4 \pi^2 \alpha_s
\frac{e}{\sqrt{2}} \frac{\delta^{a b}}{2 N_c} \, f_\rho \alpha(k_i) \int\limits_0^1 dz_i \, (z_i -\zb_i)  \, \phi(z_i) \,  \epsb  \cdot{\rm{\Qb}}(z_i,\kb,\rb,\mu_i)   \, ,
\eea
where
\beq
\label{defQ}
\Qb(z_i,\kb,\rb,\mu_i)=
 \frac{z_i \, \rb}{z_i^2\rb^2 + \mu_i^2} -
\frac{\zb_i \, \rb}{\zb_i^2\rb^2 + \mu_i^2}
 + \frac{\kb - z_i \, \rb}{(z_i\rb -\kb)^2 + \mu_i^2} - \frac{\kb-\zb_i \rb }{(\zb_i \, \rb
-\kb)^2
+ \mu_i^2}
\eq
is proportional to the impact factor of quark pair production from a
transversally polarized photon.

In the formulae (\ref{defPP}) and (\ref{defQ}) and for the rest of the paper, we denote $\mu_i^2=Q_i^2 \; z_i \; \zb_i + m^2$, where $m$ is the quark mass.
The limit $m \to 0$ is regular and we will restrict ourselves to
the light quark case, taking thus $m=0.$
Both impact factor (\ref{defPP}) and (\ref{defQ}) vanish when $\kb \to 0$ or $\rb-\kb \to 0$ due to QCD gauge invariance.

In the formulae (\ref{ifL}, \ref{ifT}), $\phi$ is the distribution
amplitude of the produced longitudinally polarized
$\rho^0-$mesons. For the case with quark $q$ of one flavour
it is defined (see, e.g. \cite{BB})
by the matrix element of the non-local, gauge invariant correlator
of quark fields on the light-cone
\begin{equation}
\label{da}
\langle 0| \bar q(x)\;\gamma^\mu\;q(-x)|\rho_L(p)= \bar{q} q \rangle = f_{\rho} \;p^\mu
\int\limits_0^1 dz \, e^{i(2z-1)(px)}\phi(z)\;,
\end{equation}
where the coupling constant is $f_\rho=216$ MeV and where the gauge links
are omitted to simplify the notation. $\phi$ is normalized to unity. The amplitudes for production of
$\rho^0$'s are obtained by noting that $|\rho^0\rangle =
1/\sqrt{2}(|\bar u u\rangle -|\bar d d \rangle )$.

Note that Eq.(\ref{da}) corresponds to the leading twist collinear distribution amplitude. Such an
object  can be used strictly speaking for asymptotically large $Q^2.$
In the phenomelogical application of sections \ref{results} and \ref{eeresults},
in order to get measurable cross-sections,
the dramatic decrease of the amplitudes with increase of  $Q_i^2,$
combined with the experimental condition of ILC project,  requires rather low values of $Q_i^2$
(of the order of 1 $\rm GeV^2$) for which subleading twist contributions could be significant. This can be taken into
account within a more phenomenological approach which incorporates intrisic $k_T$
quark distribution and which goes beyond standard QCD collinear factorization \cite{van}.
In the present paper we do not consider these effects and adhere to the collinear QCD  factorization.

Let us label the amplitudes for the scattering process (\ref{process})
through  the polarization of the incoming virtual photons as
${\cal M}_{\la_1 \la_2}.$ They can be calculated using Eqs.(\ref{M}) and Eqs.(\ref{ifL}-\ref{defQ}) supplemented by the choice of 
the transverse polarization vectors of the photons
\beq
\label{polarizationT}
\epsb^{\pm} =\frac{1}{\sqrt{2}}(\mp 1,-i) 
\eq
and the longitudinal polarization vectors (\ref{inpolL}).
 For the case $\la_1 =\la_2=0:$
\beq
\label{MCalgeneral00}
{\cal M}_{00} = i\, s\, C \, Q_1\, Q_2 \, \intfeyn d z_1 \, d z_2 \, z_1
\, \zb_1 \, \phi(z_1)\, z_2
\, \zb_2 \, \phi(z_2) {\rm M}_{00}(z_1,\, z_2)\,,
\eq
with
\beq
\label{defM00}
\hspace{-.8cm}{\rm M}_{00}(z_1,\, z_2)=\int \frac{d^2 \kb}{\kb^2 (\rb-\kb)^2} \, \rm{P_P(z_1,\kb,\rb,\mu_1)} \, \rm{P_P(z_2,-\kb,-\rb,\mu_2)}\,;
\eq
for the case $\la_2 =+,-:$
\beq
\label{MCalgeneral0i}
{\cal M}_{0\la_2} = i\, s\, \frac{C}{2} \, Q_1 \, \intfeyn d z_1 \, d z_2 \, z_1
\, \zb_1
 \, \phi(z_1)\, (z_2 - \zb_2)\, \phi(z_2) {\rm M}_{0\la_2}(z_1,\, z_2)\,,
\eq
with
\beq
\label{defM0i}
\hspace{-.8cm}{\rm M}_{0 \la_2}(z_1,\, z_2)=\int \frac{d^2 \kb}{\kb^2 (\rb-\kb)^2} \, \rm{P_P(z_1,\kb,\rb,\mu_1)} \, \rm{\Qb}(z_2,-\kb,-\rb,\mu_2) \cdot \epsb^{\la_2}\,;
\eq
for the case $\la_1 =+,-:$
\beq
\label{MCalgenerali0}
{\cal M}_{\la_1 0} = i\, s\, \frac{C}{2} \, Q_2 \, \intfeyn d z_1 \, d z_2 \, (z_1 - \zb_1)
 \, \phi(z_1)\, z_2
\, \zb_2 \, \phi(z_2) {\rm M}_{\la_1 0}(z_1,\, z_2)\,,
\eq
with
\beq
\label{defMi0}
\hspace{-.8cm}{\rm M}_{\la_1 0}(z_1,\, z_2)=\int \frac{d^2 \kb}{\kb^2 (\rb-\kb)^2} \,  \rm{\Qb}(z_1,\kb,\rb,\mu_1) \cdot \epsb^{\la_1}\, \rm{P_P(z_2,-\kb,-\rb,\mu_2)} \,.
\eq
and for the case $\la_1 =+,- \, , \la_2 =+,-:$
\beq
\label{MCalgeneralij}
{\cal M}_{\la_1 \la_2} = i\, s\, \frac{C}{4}  \, \intfeyn d z_1 \, d z_2 \, (z_1 - \zb_1)
 \, \phi(z_1)\, (z_2-\zb_2) \, \phi(z_2) {\rm M}_{\la_1 \la_2}(z_1,\, z_2)\,,
\eq
with
\beq
\label{defMij}
\hspace{-.8cm}{\rm M}_{\la_1 \la_2}(z_1,\, z_2)=\int \frac{d^2 \kb}{\kb^2 (\rb-\kb)^2} \,  \rm{\Qb}(z_1,\kb,\rb,\mu_1) \cdot \epsb^{\la_1(1)}\,\rm{\Qb}(z_2,-\kb,-\rb,\mu_2) \cdot \epsb^{\la_2(2)} \,.
\eq
Here and in the rest of this paper, we denote $C=2 \,\pi \,\frac{N_c^2-1}{N_c^2}\, \alpha_s^2 \,\alpha_{em} \, f_\rho^2 \,.$
In terms of the above amplitudes, the corresponding differential cross-sections can be expressed
in the large $s$ limit (neglecting terms of order $Q_i^2/s$) as
\beq
\label{crosssection}
\frac{d \sigma^{\gamma^*_{\la_1} \gamma^*_{\la_2} \to \rho^0_L \rho^0_L}}{dt}=\frac{|{\cal M}_{\la_1 \la_2}|^2}{16
\, \pi \,s^2 }\,
\eq
and it does not depend on s.

\section {Non-forward Born order differential cross-section for
 $\gamma^*_{L,T}\;\gamma^*_{L,T} \to \rho_L^0  \;\rho_L^0$ }
\label{cross}
\setcounter{equation}{0}

\subsection{Analytical results for $k_\perp$-integrated amplitude $M_{\la_1 \la_2}$}
\label{analytical}

In this section we summarize the results for the amplitudes $M_{\la_1 \la_2}$ obtained
after performing analytically the $k_\perp$ integrals.
Such analytic expressions give us the effective possibility of studying various kinematical limits in the variables $Q_1 ^2, \, Q_2 ^2, \,t$.
 The  $k_\perp$ integrations were done using the method of Ref.\cite{psw}
which exploits in an efficient way the scaling properties of integrals 
appearing in conformal field theories. 
The generic $k_\perp$ integral involves an integrand which corresponds to a 
box diagram with two distinct massive
propagators and two massless propagators. Because of that, the $k_\perp$ integrations result 
in long and complicate expressions. Thus, we 
discuss below 
only the general structure of the results and 
we relegate all technical details of $k_\perp$ integrations to the Appendix.

In the transverse-transverse (TT) case,
the amplitude can be expressed in term
of two projection operators in the transverse plane as follows:
\beq
\label{evaluateMij}
{\rm M}_{\la_1 \la_2}(z_1,\, z_2)= \left[a(\rb;Q_1,Q_2;z_1,z_2)  \left(\delta^{ij}-\frac{r^i r^j}{r^2}\right) + b(\rb;Q_1,Q_2;z_1,z_2) \frac{r^i r^j}{r^2} \right ] \eps^{\la_1}_i \eps^{\la_2}_j \, ,
\eq
where we denote $\rb^2 =r^2.$

Combining (\ref{MCalgeneralij}) and (\ref{evaluateMij}), and using $| M_{++}|^2 = | M_{--}|^2 ,$ one gets
in the case of two photons with the same polarization :
\beqa
\label{M2Calgeneralpp}
&&|{\cal M}_{++}|^2 =  |{\cal M}_{--}|^2 = s^2\, \frac{C^2}{64} \\
&& \hspace{-.3cm}\times \,\left| \intfeyn d z_1 \, d z_2 \, (z_1 - \zb_1)
\, \phi(z_1)\, (z_2-\zb_2) \,\phi(z_2) \, \left(b(\rb;Q_1,Q_2;z_1,z_2) - a(\rb;Q_1,Q_2;z_1,z_2)\right) \right|^2 \,,\nonumber
\eqa
and analogously for different polarizations :
\beqa
\label{MCalgeneralpm}
\hspace{-1cm}&&|{\cal M}_{+-}|^2 =  \, s^2\, \frac{C^2}{64} \\
\hspace{-1cm}&&\times  \,\left| \intfeyn d z_1 \, d z_2 \, (z_1 - \zb_1)
\, \phi(z_1)\, (z_2-\zb_2) \,\phi(z_2) \, \left(b(\rb;Q_1,Q_2;z_1,z_2) + a(\rb;Q_1,Q_2;z_1,z_2)\right) \right|^2 \nonumber \,.
\eqa
For the longitudinal-transverse (LT) case, restoring the dependency over all variables, one defines from (\ref{defM0i}) and (\ref{defMi0})  the scalar function $f$
\beq
\label{deffM0i}
{\rm M}_{0 \la}(\rb;Q_1,Q_2;z_1,\, z_2)= f(\rb;Q_1,Q_2;z_1,\, z_2) \, \rb \cdot \epsb^{\la}\,,
\eq
or equivalently
\beq
\label{deffM0ibis}
{\rm M}_{\la 0}(\rb;Q_1,Q_2;z_1,\, z_2)= f(\rb;Q_2,Q_1;z_2,\, z_1) \, \rb \cdot \epsb^{\la} \,,
\eq
which leads to
\beq
\label{M2Calgeneral0i}
|{\cal M}_{0+}|^2 =  |{\cal M}_{0-}|^2=s^2\, \frac{C^2}{8} \, Q_1^2\, \rb^2 \,\left| \intfeyn d z_1 \, d z_2 \, z_1
\, \zb_1
 \, \phi(z_1)\, (z_2 - \zb_2)\, \phi(z_2)  f(\rb;Q_1,Q_2;z_1,\, z_2)\right|^2\!.
\eq
and analogously for the transverse-longitudinal (TL) case
\beq
\label{M2Calgenerali0}
|{\cal M}_{+0}|^2 =  |{\cal M}_{-0}|^2 =s^2\, \frac{C^2}{8} \, Q_2^2\, \rb^2 \,\left| \intfeyn d z_1 \, d z_2 \, z_2
\, \zb_2
 \, \phi(z_2)\, (z_1 - \zb_1)\, \phi(z_1)  f(\rb;Q_2,Q_1;z_2,\, z_1)\right|^2 \!.
\eq

The expressions of $a(\rb;Q_1,Q_2;z_1,z_2)$, $b(\rb;Q_1,Q_2;z_1,z_2)$ and $f(\rb;Q_1,Q_2;z_1,\, z_2)$  presented as  combinations of 
{\it finite} standard integrals are given in the Appendix.

For the longitudinal-longitudinal (LL) case, it turned out  \cite{psw} that (\ref{MCalgeneral00}) can be effectively replaced by $\tilde{M}(z_1,\, z_2)$ whose integral over $z_{1,2}$ with symmetrical DA gives
the same result.  $\tilde{M}(z_1,\, z_2)$
 reads
\bea
\label{Mcompact2}
&&\!\!\!\!\! \tilde{M}_{00}(z_1,\, z_2) = - \left( \frac{1}{z_1^2 \rb^2 +\mu_1^2}
+\frac{1}{\zb_1^2 \rb^2 +\mu_1^2}\right) J_{3 \mu_2}(z_2) -\left( \frac{1}{z_2^2 \rb^2 +\mu_2^2}
+\frac{1}{\zb_2^2 \rb^2 +\mu_2^2}\right) J_{3 \mu_1}(z_1)
\nonumber \\
&&+J_{4 \mu_1 \mu_2}(z_1,\, z_2) + J_{4 \mu_1 \mu_2}(\zb_1,\, z_2)\;.
\eea
$J_{3 \mu}$ and $J_{4 \mu_1 \mu_2}$ are two dimensional integrals with
respectively 3 propagators (1 massive) and 4 propagators (2 massive,
with different masses), they are both IR and UV finite. Their expressions are given in the Appendix.

Due  to the collinear conformal subgroup $SL(2, R)$ invariance \cite{Braun}, the $\rho^0_L$ distribution amplitude has an expansion in terms of Gegenbauer polynomials
of even order which reads
\beq
\label{asymptDA}
\phi(z)=6 z (1-z)\,(1+ \sum_{n=1}^{\infty} a_{2 \, n} C^{3/2}_{2\,n} (2 z-1))\,.
\eq
Except for a short discussion in section \ref{eeresults}, we restrict ourselves to the asymptotical 
distribution amplitude corresponding to $a_{2\,n}=0.$

To complete the evaluation of the amplitude ${\cal M},$ one needs to integrate
over the quark momentum fractions  $z_1$ and $z_2$ in the $\rho$ mesons.
For arbitrary values of $t,$ it seems not possible to
perform the $z_1$ and $z_2$ integrations analytically.
We thus do them numerically.
We observe the absence of end-point singularity when $z_{1(2)} \to 0$ or
$z_{1(2)} \to 1.$ Indeed, for the longitudinal polarizations involving $P_P$ as defined in Eq.(\ref{defPP}),
the $z$  divergency of type
$1/z,$ $1/(1-z)$  is compensated by the $z\zb$ factor when $z \to 0,\, 1,$ while for transverse polarizations, involving   $\underline Q$ as defined in
Eq.(\ref{defQ}), there is no singularity since $\underline Q$ is itself regular.

For the special case $t=t_{min}$ (where only the LL amplitude is non-vanishing), which will be useful in the discussion of sections  \ref{results}
and \ref{eeresults}, the integration over $z_i$ can be performed analytically, with the result
\cite{psw}
\bea
\label{resultM00min}
&&\hspace{-1.1cm}{\cal M}_{00}= -i s \, \frac{N_c^2-1}{N_c^2} \, \alpha_s^2 \, \alpha_{em}  \, f_\rho^2 \, \frac{9 \pi^2}{2} \, \frac{1}{Q_1^2
  Q_2^2}  \left[6 \, \left(R + \frac{1}{R}\right) \ln^2 R  + \,12\,
\left(R-\frac{1}{R}\right) \ln R \right. \nonumber\\
&&\hspace{-1cm}\left. +\, 12 \,  \left(R + \frac{1}{R}\right)\,+\,
\left(3 \,R^2 +\, 2\,+\frac{3}{R^2}\right) \left(\,\ln\,(1-R)\,\ln^2 R-\ln \,(R+1)\,\ln^2 R 
  \nonumber \right. \right.\\
&&\hspace{-1cm}\left. \left.-\,2 \,{\rm Li_2}\,(-R)\,
    \ln R \,+\, 2\,{\rm Li_2}\,(R)\, \ln R \,+\,2\,{\rm Li_3}\,(-R)\,-\,2\, {\rm Li_3}\,(R)\right)\right]\,,
\eea
where $R=Q_1/Q_2\,.$ When $Q_1=Q_2,$ the expression  (\ref{resultM00min}) simplifies to
\beq
\label{bornggR1} 
{\cal M}_{00} = is \frac{N_c^2-1}{N_c^2}\,  \alpha_s^2 \alpha_{em} f_\rho^2 
\, \frac{9\pi^2}{Q^4}  (14\zeta(3)-12) \,.
\eq 
  
\subsection{Results for differential cross-section}
\label{results}

The formulae for ${\rm M}_{\la_1 \la_2}$ obtained in sec.\ref{cross} permit us to 
evaluate the magnitudes of  cross-sections (\ref{crosssection}) of the diffractive double rho production 
for different helicities of virtual photons. 
In our estimates we use as a strong coupling constant  the 
three-loop running  $\alpha_S(Q_1 Q_2)$ with $\Lambda_{\overline{MS}}^{(4)}= 305 \,$MeV
(see, e.g. \cite{bethke}).\footnote{Running of $\alpha_S$ is in principle a subleading effect with respect to our treatment.
Nevertheless, numerically, as we discuss in sec.\ref{eeresults}, the dependence of our predictions 
for the rates in $e^+e^-$ scattering on a choice of $\alpha_s$ is  negligible at Born order, but is more subtle when LO BFKL corrections are taken into account.}
 
\begin{figure}
\begin{picture}(800,600)
\put(0,400){\epsfxsize=\wid{\centerline{\epsfbox{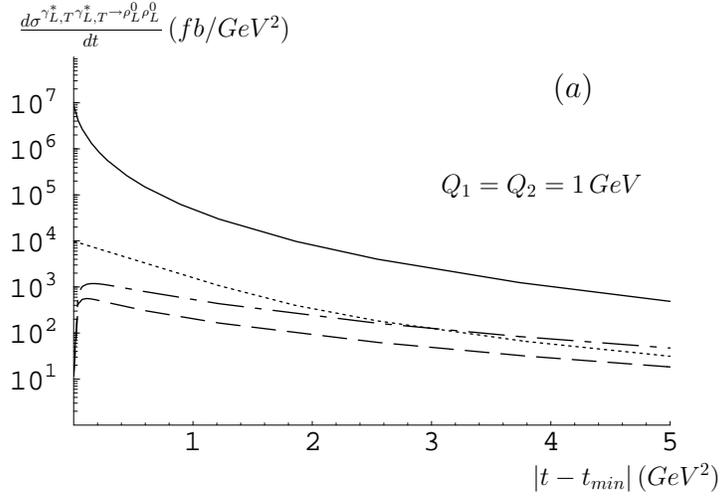}}}}
\put(0,200){\epsfxsize=\wid{\centerline{\epsfbox{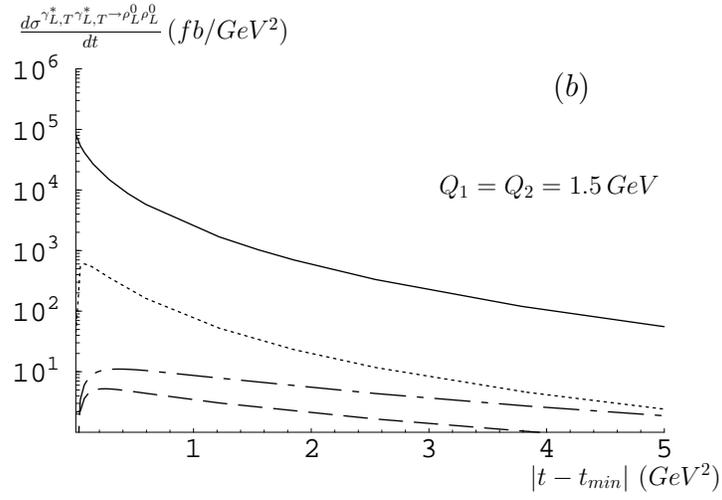}}}}
\put(0,0){\epsfxsize=\wid{\centerline{\epsfbox{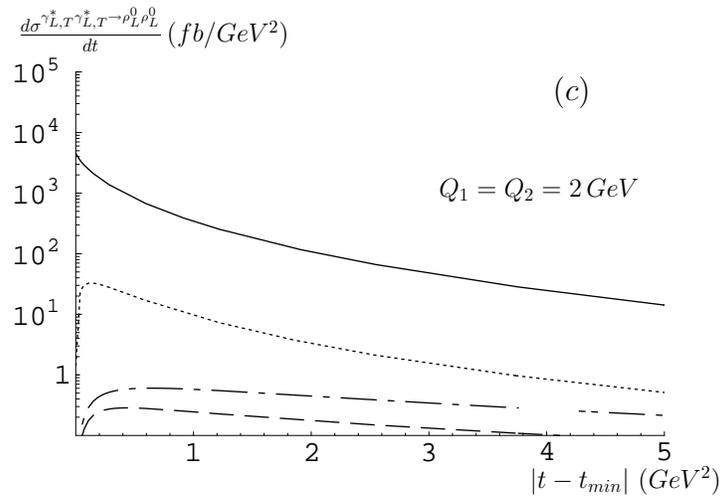}}}}
\put(300,550){$(a)$}
\put(300,350){$(b)$}
\put(300,150){$(c)$}
\end{picture}
\caption{\small Differential cross-sections for the process $\gamma^*_{L,T}\;\gamma^*_{L,T} \to \rho_L^0  \;\rho_L^0$. The solid curve corresponds to the  $\gamma^*_L \gamma^*_L$ mode, the dotted one to the $\gamma^*_L\gamma^*_T$ mode,  the dashed and the dashed-dotted ones to the $\gamma^*_T \gamma^*_{T'}$ modes with respectively  the same $T=T'$ and  different $T \neq T'$ transverse polarizations.
The different figures (a), (b), (c) correspond to different values of $Q_1=Q_2.$}
\label{FigQGeV}
\end{figure}

In Fig.\ref{FigQGeV} we display
the $t$-dependence
of the different
$\gamma^*_{L,T}\;\gamma^*_{L,T} \to \rho_L^0  \;\rho_L^0$ differential cross-sections for various values of $Q=Q_1=Q_2.$


We first note  the strong decrease of all the cross-sections when $Q_{1,2}^2$ increase. For LL, this follows from an obvious dimensional analysis, since 
$${\cal M}_{LL} \, \propto  \, \frac{s \, f_\rho^2}{Q^4}\,$$
(for $Q_1=Q_2=Q$), in agreement with (\ref{bornggR1}).

Secondly, all the differential cross-sections which involve at least one transverse
photon vanish when $t=t_{min}.$ It is due to the vanishing 
of the function $\underline Q$ for $\underline r=0$
(see (\ref{defQ})). Physically,
this fact is related to the $s$-channel helicity conservation at $t=t_{min}$.
Indeed, since the $t$-channel gluons carry non-sense polarizations,
helicity conservation occurs separately in each impact factor.

\begin{figure}[h]
\begin{picture}(800,380)
\put(-120,200){\epsfxsize=\wids{\centerline{\epsfbox{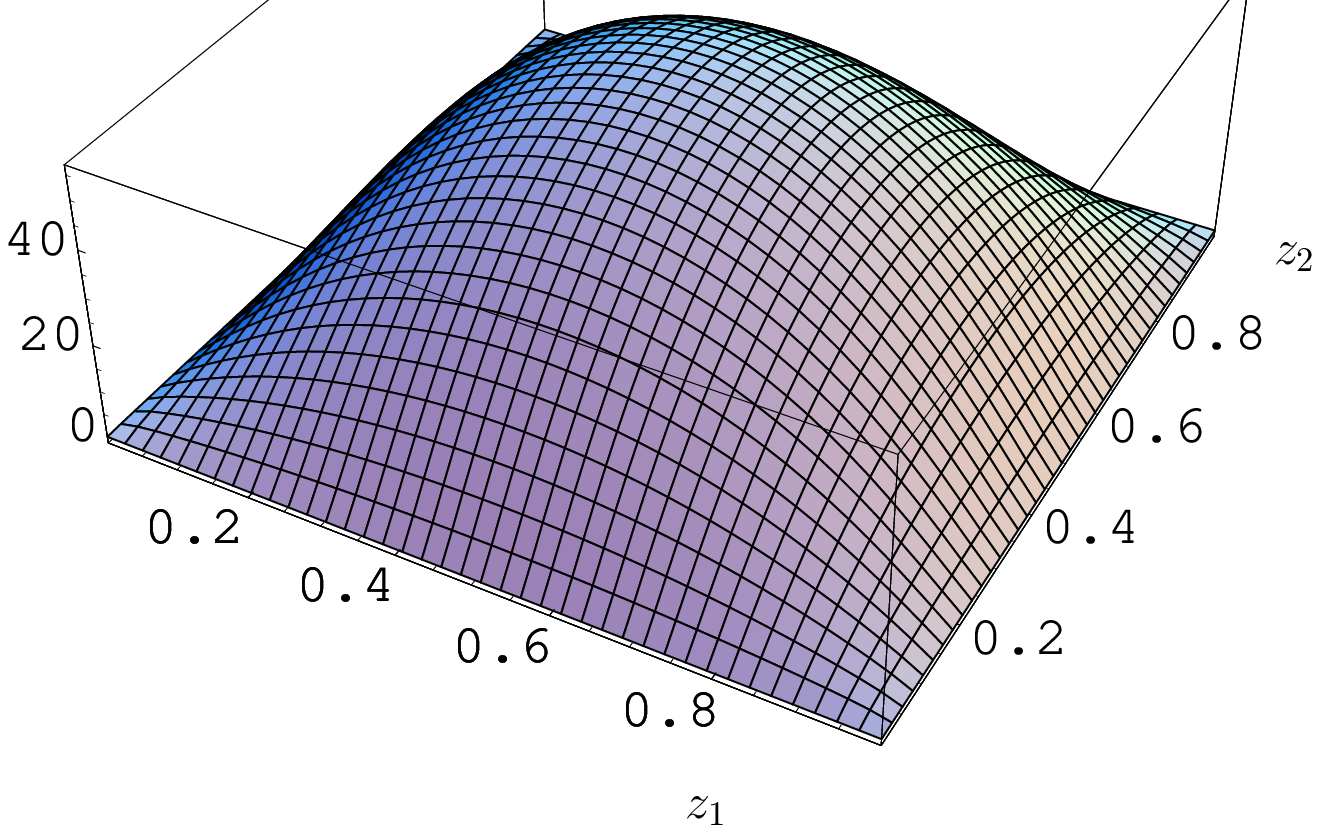}}}}
\put(120,200){\epsfxsize=\wids{\centerline{\epsfbox{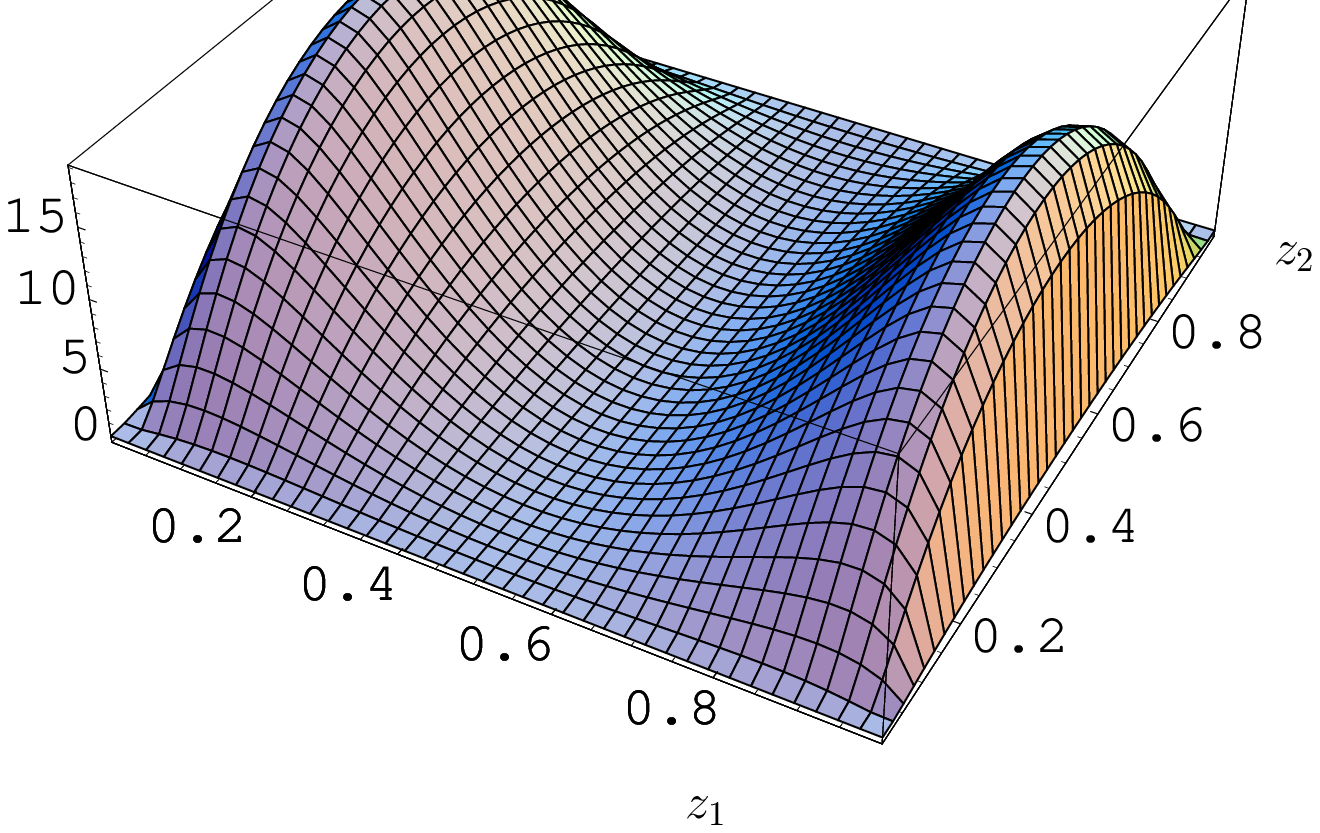}}}}
\put(-120,0){\epsfxsize=\wids{\centerline{\epsfbox{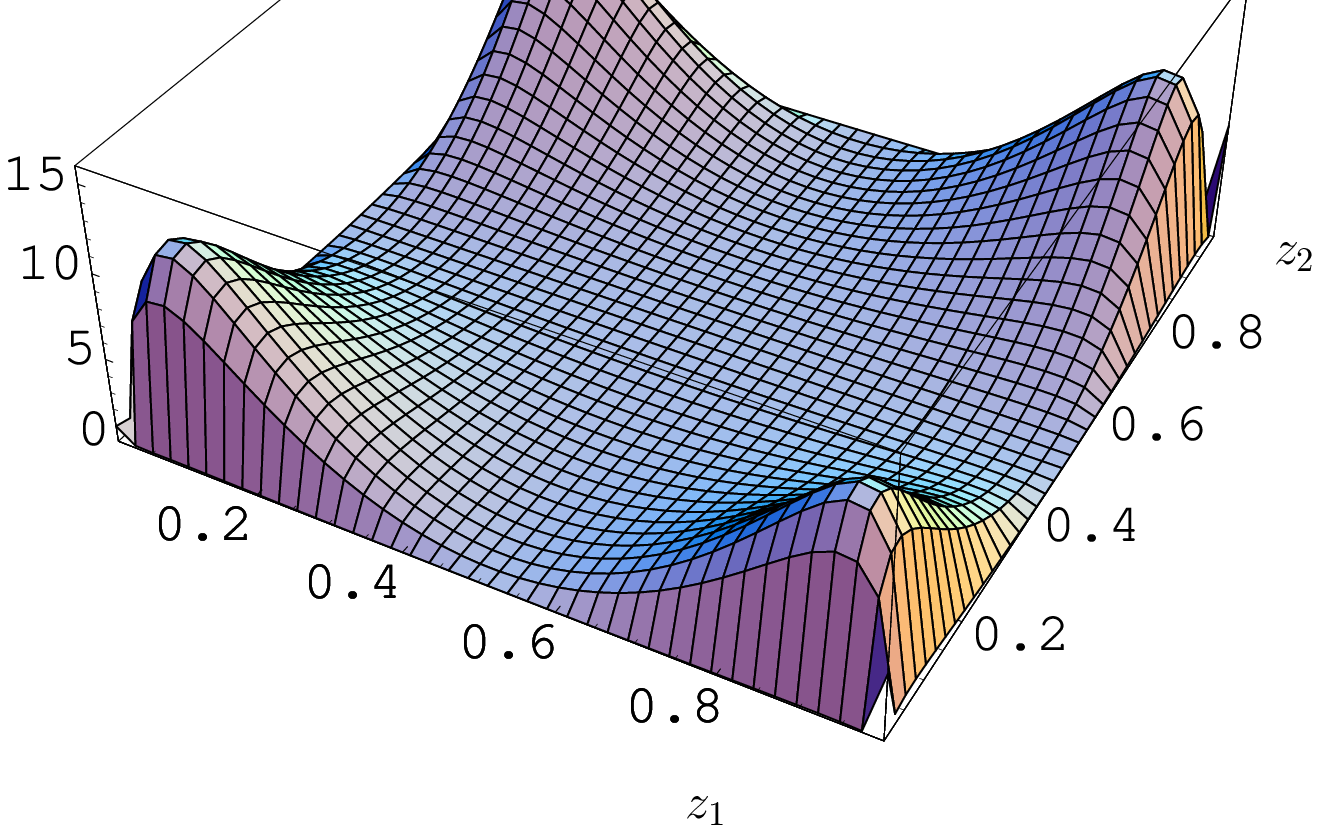}}}}
\put(120,0){\epsfxsize=\wids{\centerline{\epsfbox{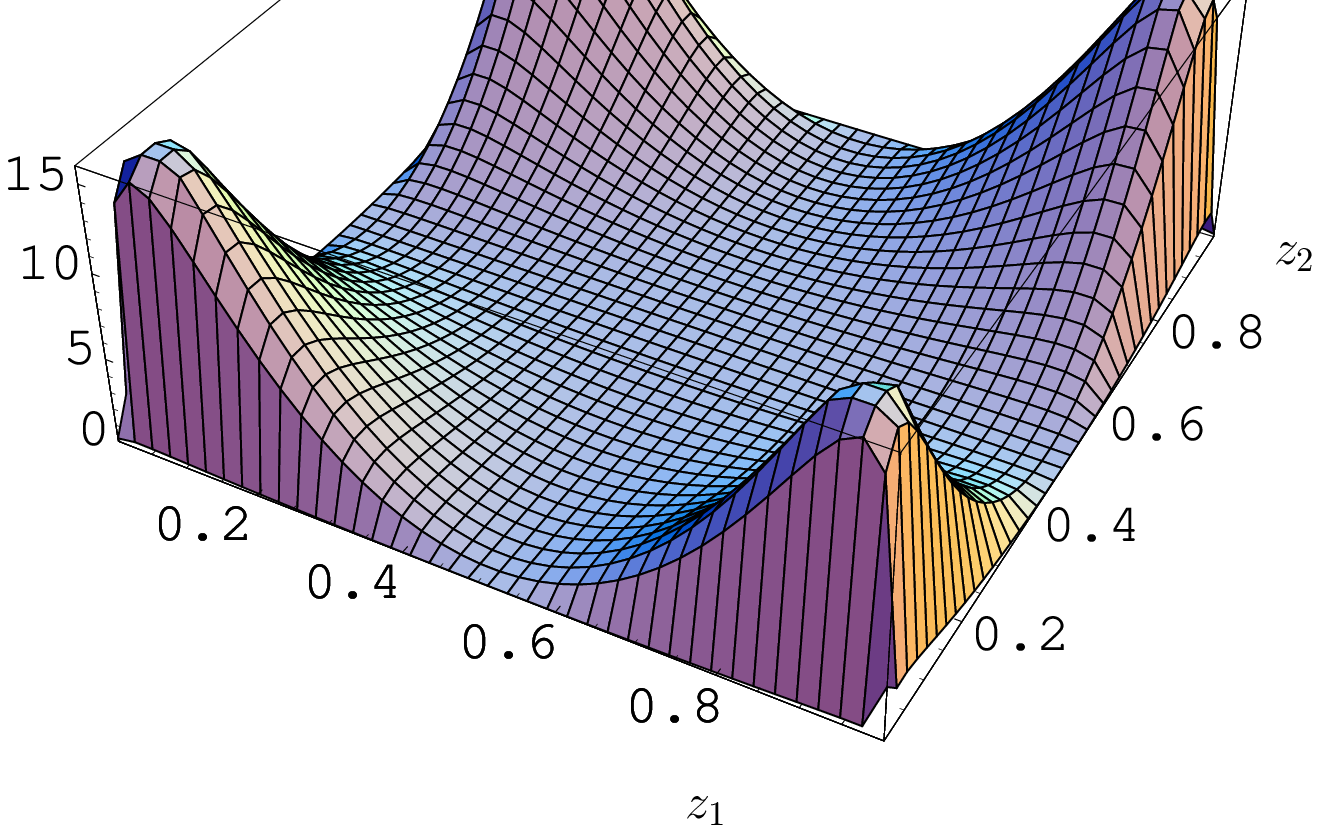}}}}
\put(20,350){$\overline{\rm M}_{00}$}
\put(260,350){$-\overline{\rm M}_{+0}$}
\put(20,150){$\overline{\rm M}_{++}$}
\put(260,150){$-\overline{\rm M}_{+-}$}
\end{picture}
\caption{\small Shape of the amplitudes $\overline{\rm M}_{00},$ $-\overline{\rm M}_{+0},$  $\overline{\rm  M}_{++},$ $-\overline{\rm  M}_{+-}$ as functions of $z_1$ and $z_2,$ for $-t=0.16 \,\rm  GeV^2$ and
$Q_1=Q_2=1 \, \rm GeV.$}
\label{Figshape1}
\end{figure}
In Fig.\ref{Figshape1},
we show the shape of the integrands ${\overline{\rm M}}_{\lambda_1, \la_2}$ of the various amplitudes ${\cal M}_{\la_1 \la_2}$ as a function of $z_1$ and $z_2$, as they appear in formulas (\ref{MCalgeneral0i}, \ref{MCalgenerali0} and \ref{MCalgeneralij}):
\beq
\label{Mbar00}
{ \overline{\rm M}}_{0 0} = z_1
\, \zb_1 \, \phi(z_1)\, z_2
\, \zb_2 \, \phi(z_2) \,{\rm M}_{00}(z_1,\, z_2)\,,
\eq
and for $\la_i=+,-$
\beqa
\label{Mbari0}
{ \overline{\rm M}}_{\la_1 0} &=& (z_1 - \zb_1)
 \, \phi(z_1)\, z_2
\, \zb_2 \, \phi(z_2)\, {\rm M}_{\la_1 0}(z_1,\, z_2)\,,\\
\label{Mbarij}
{ \overline{\rm M}}_{\la_1 \la_2} &=&  (z_1 - \zb_1)
 \, \phi(z_1)\, (z_2-\zb_2) \, \phi(z_2) \,{\rm M}_{\la_1 \la_2}(z_1,\, z_2)\,.
\eqa
${\rm M}_{\la_1 \la_2}(z_1,\, z_2)$ is symetric under $(z_i\lr \zb_i)$ for a longitudinal polarization $\la_i=0$ (cf. \ref{defPP}) and antisymetric under $(z_i\lr \zb_i)$ for a transverse polarization $\la_i=+,-$ (cf. \ref{defQ}); thus the factors  $z_i \, \zb_i$ for $\la_i=0$ and $z_i - \zb_i$ for $\la_i=+,-$ ensure the symmetry of $ \overline{\rm M}_{\la_1 \la_2}$ under $(z_i\lr \zb_i)$ as we can see on Fig.\ref{Figshape1}. Because of the $\rho^0_L$ mesons distribution amplitudes $\phi(z_i)$, $\overline{\rm M}_{\la_1 \la_2}(z_1,\, z_2)$ vanishes for any polarization in the end-point region. Consequently the case of a transverse polarization vanishes in the central region $z_i=\zb_i=1/2$ and also in the end-point region $z_i$ close to 0 or 1, so that it restricts the available $z_i$ phase-space and reduces the resulting differential cross-section, in agreement with the dominance of longitudinal photons (helicity conservation) in the process $\gamma^*_{L,T}\;\gamma^*_{L,T} \to \rho_L^0  \;\rho_L^0$.

 The amplitude involving at least one transverse photon
has a maximum at low $-t$ value with respect to $Q_1 Q_2$. The Fig.\ref{Figshape1} corresponds to $-t=0.16$ $ \rm GeV^2$ which is a typical value for the 
region where the cross-sections with transverse photons in Fig.\ref{FigQGeV} are 
maximal.

\begin{figure}[h]
\begin{picture}(800,170)
\put(-120,0){\epsfxsize=\wids{\centerline{\epsfbox{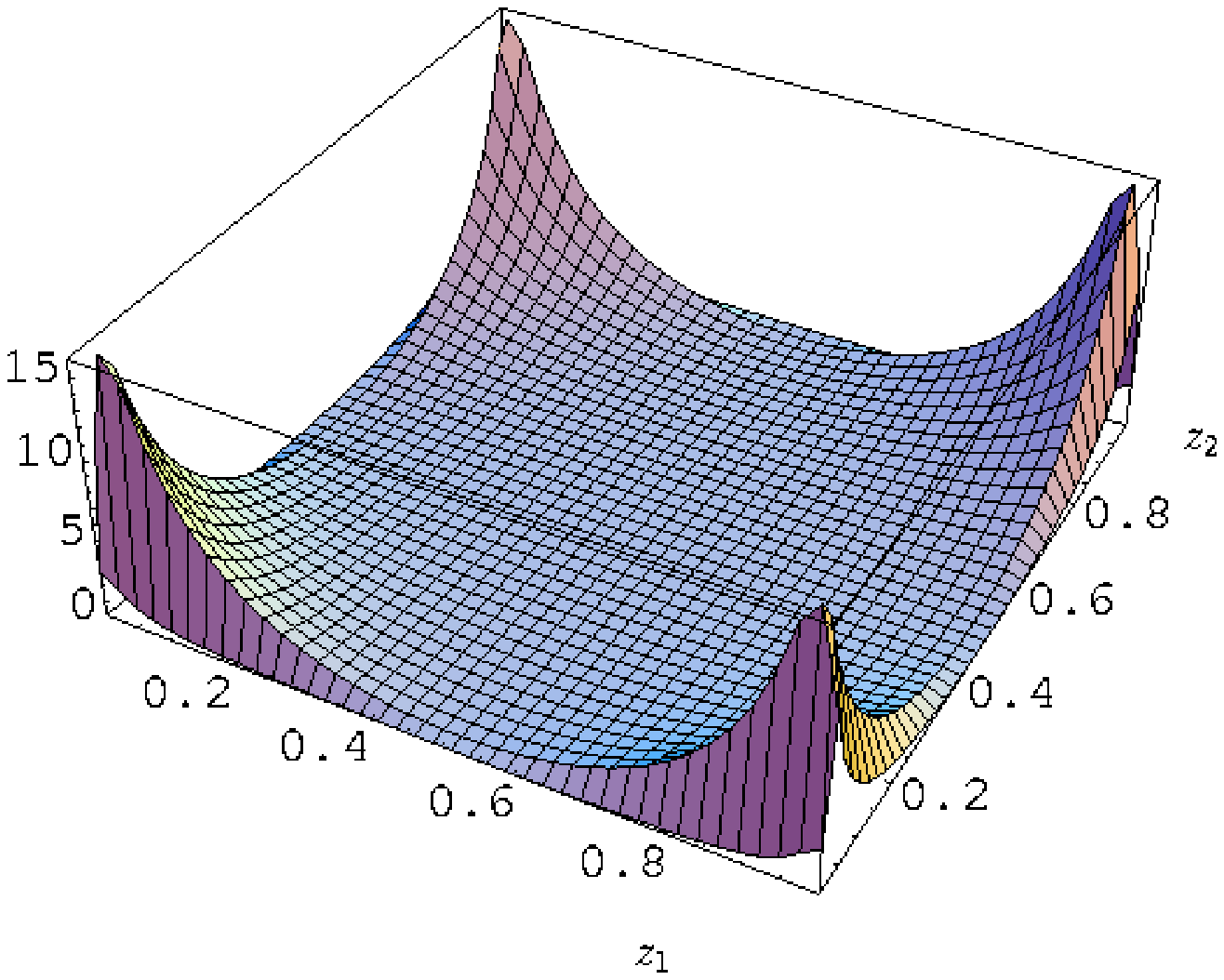}}}}
\put(120,0){\epsfxsize=\wids{\centerline{\epsfbox{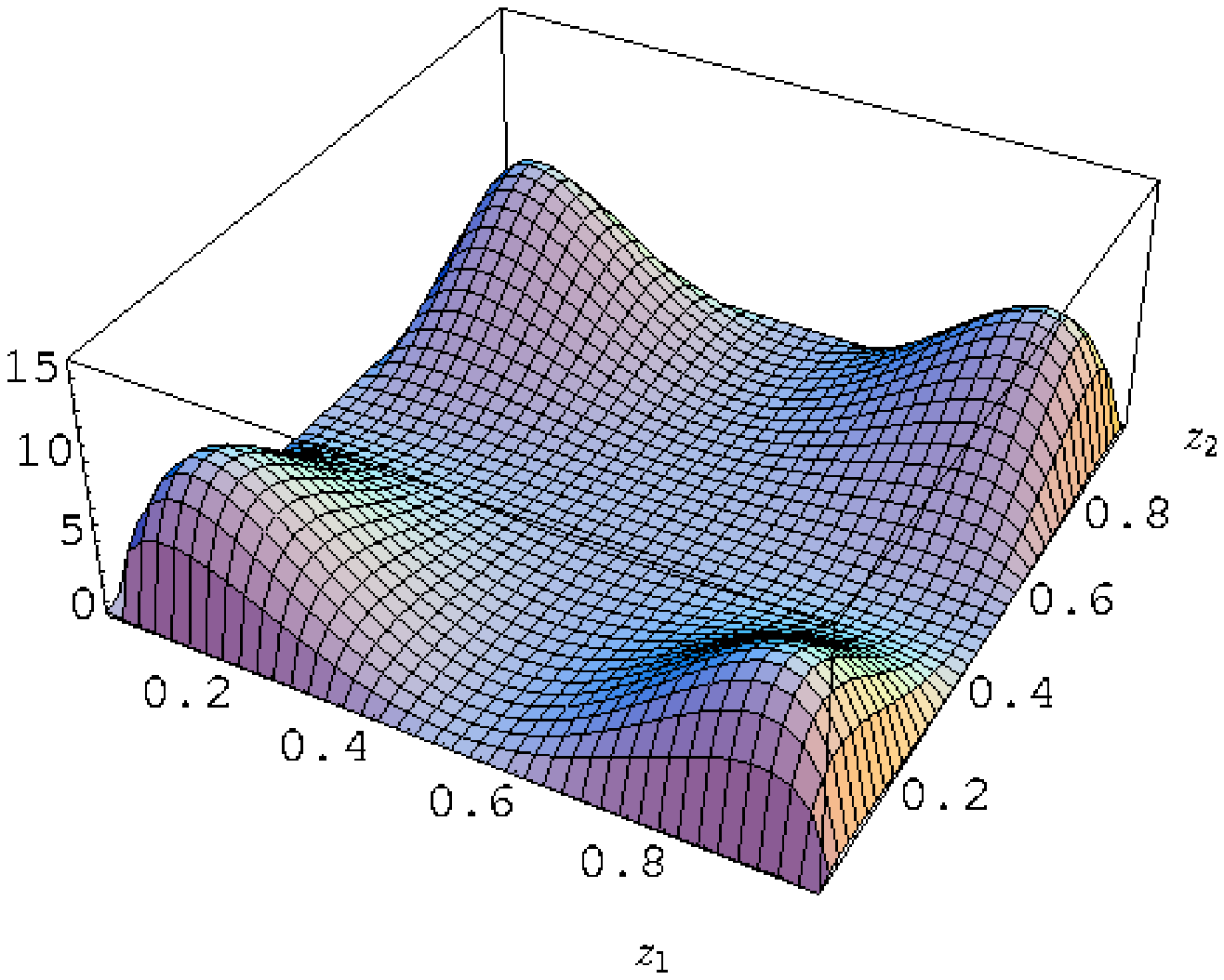}}}}
\end{picture}
\caption{\small Shape of the amplitude $-\overline{\rm M}_{+-},$ for $-t=0.01 \, \rm GeV^2$ (left) and $-t=0.8 \, \rm GeV^2$  (right), with
$Q_1=Q_2=1 \, \rm GeV.$}
\label{Figshape2}
\end{figure}
A peculiarly characteristic shape appears in the amplitudes with two 
transverse photons, as shown in bottom panels of Fig.\ref{Figshape1}.
When the value of t changes towards $t_{min}$ the peaks become 
very narrow, as shown in the left panel in Fig.\ref{Figshape2} for $\overline{\rm M}_{+-}$. For $t$ very 
close to $t_{min}$ they are practically concentrated only on the boundary which leads to 
the vanishing of the amplitude. On the other hand, when the value of t
increases and leaves the maximum of cross-sections  the peaks in Fig.\ref{Figshape1} 
decrease and spread, as shown for $\overline{\rm M}_{+-}$ in the right panel of Fig.\ref{Figshape2}.
 
\begin{figure}[h]
\epsfxsize=\wid{\centerline{\epsfbox{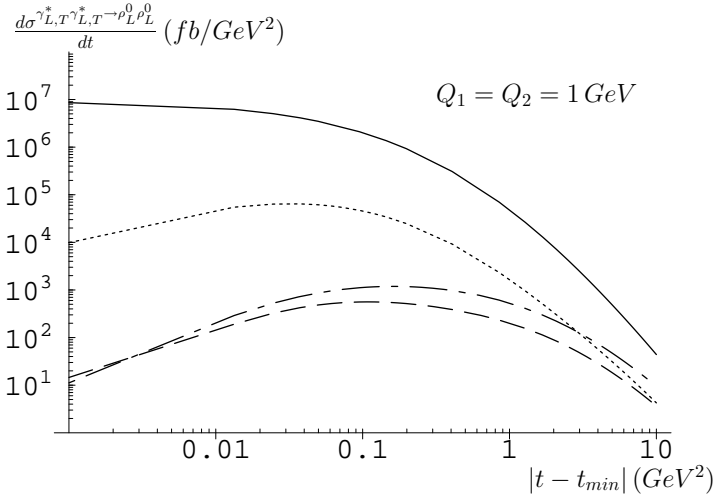}}}
\caption{\small Differential cross-sections for the process $\gamma^*_{L,T}\;\gamma^*_{L,T} \to \rho_L^0  \;\rho_L^0,$ for small value of $t$.  The solid curve corresponds to the  $\gamma^*_L \gamma^*_L$ mode, the dotted one to the $\gamma^*_L\gamma^*_T$ mode,  the dashed and the dashed-dotted ones to the
 $\gamma^*_T \gamma^*_{T'}$ modes with respectively  the same $T=T'$ and  different $T \neq T'$
transverse polarizations,
for
$Q_1=Q_2=1 \, \rm GeV.$}
\label{FigDiffloglog}
\end{figure}

In the case of LT polarizations, the shape of the amplitude $\overline{\rm M}_{+0}$, which contains  only one factor $(z_i-\bar z_i)$, is shown in the right upper
panel of Fig.\ref{Figshape1}. 
Its comparison  with the upper left panel of Fig.\ref{Figshape1},
showing the shape of the $\overline{\rm M}_{00}$ amplitude, leads to the conclusion that  
$\overline{\rm M}_{+0}$ shares some properties with $\overline{\rm M}_{+-}$ and $\overline{\rm M}_{00}$. In particular, the
presence of a transverse polarization leads to  the vanishing of $\overline{\rm M}_{+0}$ at $t=t_{min}$. On the other hand, the presence of a longitudinal polarization 
increases the cross-section at small values of $t.$ As a consequence of the
competition of these two mechanisms, the maximum of the 
cross-section determined by $\overline{\rm M}_{+0}$ is located closer to $t_{min}$ than in the 
case of the cross-section given by $\overline{\rm M}_{+-}$. This is illustrated in 
Fig.\ref{FigDiffloglog} which 
shows the t-dependence of the various differential cross-sections in 
log-log scale.

\begin{figure}[h]
\epsfxsize=\widl{\centerline{\epsfbox{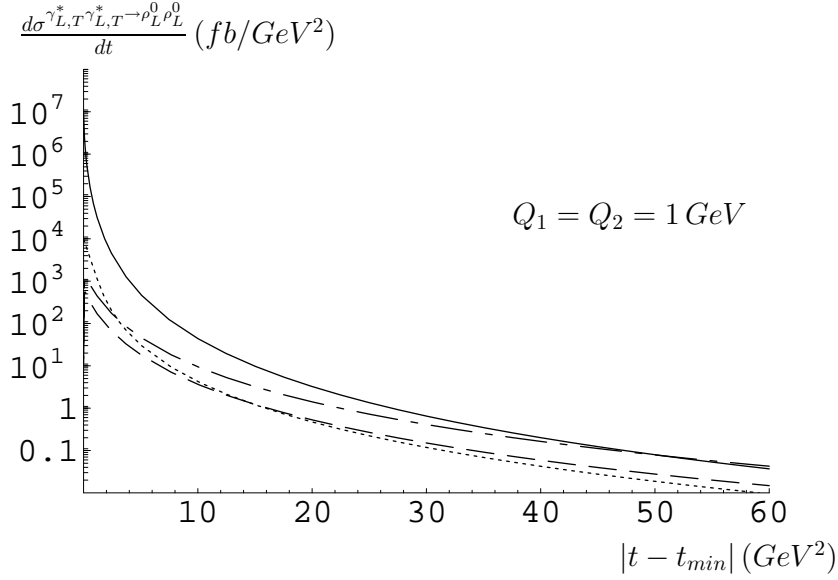}}}
\caption{\small Differential cross-sections for the process $\gamma^*_{L,T}\;\gamma^*_{L,T} \to \rho_L^0  \;\rho_L^0,$ up to asymptotically large $t.$  The solid curve corresponds to the  $\gamma^*_L \gamma^*_L$ mode, the dotted one to the $\gamma^*_L\gamma^*_T$ mode,  the dashed and the dashed-dotted ones to the 
$\gamma^*_T \gamma^*_{T'}$ modes with respectively  the same $T=T'$ and  different $T \neq T'$
transverse polarizations,
for
$Q_1=Q_2=1 \, \rm GeV.$}
\label{Figsigmalarget}
\end{figure}

Third, in  Fig.\ref{Figsigmalarget}, we display the $t$-dependence of the 
$\gamma^*_{L,T}\;\gamma^*_{L,T} \to \rho_L^0  \;\rho_L^0$ differential cross-sections for  $Q=Q_1=Q_2=1$ GeV up to values of $-t$ much larger than photon virtualities $Q_i$,  where 
$t$ plays the role of the dominant hard scale in our process. Of course, in such a
kinematical region the cross-section are strongly suppressed in comparison 
with the small $t$ one. Nevertheless, Fig.\ref{Figsigmalarget} illustrates the expected fact 
that the hierarchy of cross-sections is different in two regions: at large $t,$  the $\gamma^*_T\;\gamma^*_T \to \rho_L^0  \;\rho_L^0$ cross-section dominates over the one of $\gamma^*_L\;\gamma^*_L \to \rho_L^0  \;\rho_L^0$, which is the dominant cross-section at small $t,$
since the virtual photons are almost on shell with respect to the large scale given by $t$.

To conclude this subsection, we note that all the above cross-sections are strongly peaked in the  forward cone. The phenomenological predictions obtained in the
region of the forward cone will practically dictate the general trends of the integrated cross-sections.
This fact is less dangerous than for the real photon case since the virtual
 photon is not in the direction of the beam, and thus the outgoing $\rho$ mesons can be tagged.
 The only difficulty has to do with the tagging of the outgoing lepton, since the cross-section is dominated by small
(hard) values of  $Q_{1,2}^2.$ In this section we did not modify cross-sections
by taking into account the virtual photon fluxes, which would amplify 
both, the dominance of small $Q^2$ region as well as the small $y_i$ domain,
characteristic for very forward outgoing leptons. This is discussed in section \ref{crossee}. In particular,
it will be shown that 
the differential cross-sections are experimentaly visible and seems to be sufficient for the
$t-$dependence to be measured up to a few ${\rm GeV^2.}$

Note also that at this level of calculation there is no $s$-dependence of the
cross-section. It will appear after taking into account  triggering
effects and/or BFKL evolution.


\subsection{Quark exchange contribution to the cross-section}
\label{subdominant}

\begin{figure}[htp]
\begin{eqnarray*}
\psfrag{q1}[cc][cc]{$q_1$}
\psfrag{q2}[cc][cc]{$q_2$}
\psfrag{l1}[cc][cc]{}
\psfrag{l1t}[cc][cc]{}
\psfrag{l2}[cc][cc]{}
\psfrag{l2t}[cc][cc]{}
\psfrag{n}[cc][cc]{}
\begin{array}{cccccccc}
\raisebox{-0.44 \totalheight}{\epsfig{file=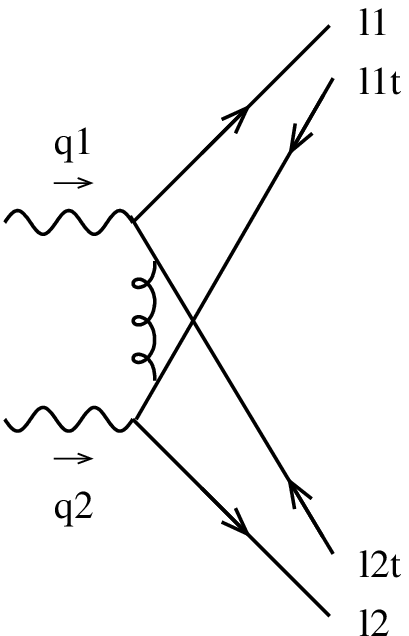,width=\scalingb}}&+&\raisebox{-0.44 \totalheight}{\epsfig{file=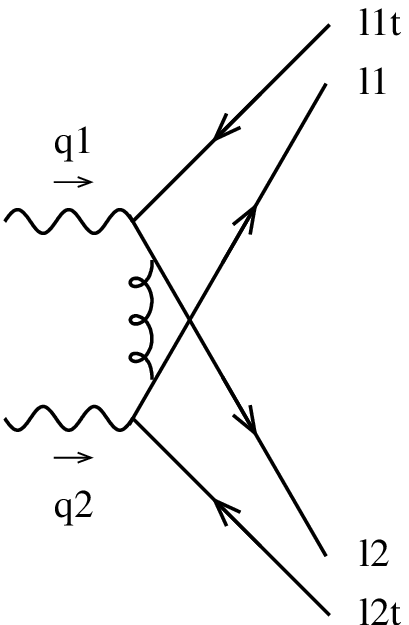,width=\scalingb}} &+&\raisebox{-0.44 \totalheight}{\epsfig{file=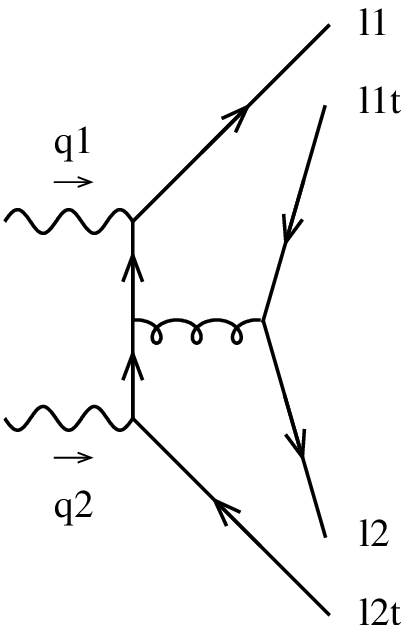,width=\scalingb}}&+&\raisebox{-0.44 \totalheight}{\epsfig{file=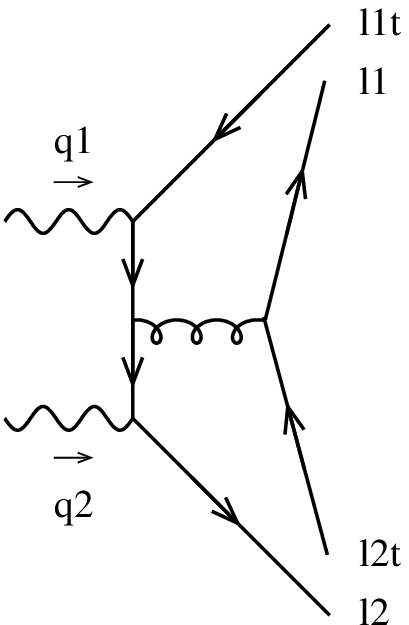,width=\scalingb}}
\end{array}
\end{eqnarray*}
\caption{\small Feynman diagrams contributing to $M_H$
in the case of
 longitudinally polarized virtual photons. \label{FiglongDiagrams}}
\end{figure}

The process (\ref{process}) described above  involves gluon exchanges which dominate at high 
energies. However, at lower energy, the process can be described by double quark
exchange. This was investigated in \cite{gdatda}, in the case $t=t_{min}.$
Fig.\ref{FiglongDiagrams} shows in particular the diagrams which contributes to the amplitude $M_H$ (see Fig.\ref{FigCollinear}) for the process $\gamma^*_L(q_1)\;\gamma^*_L(q_2) \to \rho_L(k_1)  \;\rho_L(k_2)\,.$
Using the Eqs.(12) and (13) of  \cite{gdatda} together with the asymptotical $\rho_L^0$ distribution amplitude (\ref{asymptDA})  one obtains the scattering amplitude for the photons 
longitudinally polarized 

\beq
\label{MLLcollinear} {\cal
M}^{q \bar{q}}_{00}= - \, 40\, \pi^2  \,  \frac{N_c^2-1}{N_c^2} \,
\frac{ \alpha_s \, \alpha_{em}  \, f_\rho^2 }{s} \left [ 1 +
\frac{\left( 1 +  \frac{Q_1^2}{s} +
     2  \frac{Q_1^2}{s} \frac{  \ln \frac{Q_1^2}{s}}{1 - \frac{Q_1^ 2}{s}} \right) \,
\left( 1 +  \frac{Q_2^2}{s} +2  \frac{Q_2^2}{s} \frac{  \ln \frac{Q_2^2}{s}}{1 - \frac{Q_2^ 2}{s}}
       \right)}
{\left( 1 -
       \frac{Q_1^2}{s}
       \right) \,
     \left( 1 -
       \frac{Q_2^2}{s}
       \right) } \right ]
\eq and for the transversally polarized photons 

\beqa
 \label{MTTcollinear}
{\cal M}^{q \bar{q}}_{TT} &=& {\cal M}^{q \bar{q}}_{++}+ {\cal M}^{q \bar{q}}_{--} \\
&=&- \, 40\, \pi^2  \,  \frac{N_c^2-1}{N_c^2} \, \frac{ \alpha_s
\, \alpha_{em}  \, f_\rho^2 }{s} \frac{
         \left( \frac{7}{2} +
       \frac{2\,
          \left( 1 +
            \frac{Q_1^
            2}{s} \right) \,
          \ln \frac{Q_1^2}{s}}{1 -
          \frac{Q_1^2}
           {s}} \right) \,
     \left( \frac{7}{2} +
       \frac{2\,
          \left( 1 +
            \frac{Q_2^
            2}{s} \right) \,
          \ln \frac{Q_2^2}{s}}{1 -
          \frac{Q_2^2}
           {s}} \right) -\frac{1}{4}}{\left(
      1 - \frac{Q_1^2}
       {s} \right) \,
    \left( 1 -
      \frac{Q_2^2}{s}
      \right) } .
\nonumber
 \eqa 
In the large $s$ limit, one respectively gets
\beq
\label{MLLcollinearS} {\cal M}^{q \bar{q}}_{00} \simeq - \, 80\,
\pi^2  \,  \frac{N_c^2-1}{N_c^2} \, \frac{ \alpha_s \, \alpha_{em}
\, f_\rho^2 }{ Q_1 \, Q_2}
\eq 
 and
\beqa \label{MTTcollinearS}
{\cal M}^{q \bar{q}}_{TT} 
&\simeq&- \, 40\, \pi^2  \,  \frac{N_c^2-1}{N_c^2} \, \frac{ \alpha_s \, \alpha_{em}  \, f_\rho^2 }{s}
\left( 4  \ln \frac{Q_1^2}{s} \, \ln \frac{Q_2^2}{s}+ 14 \, \ln \frac{Q_1 \, Q_2}{s} +12 \right)  \\
&=& - \, 40\, \pi^2  \,  \frac{N_c^2-1}{N_c^2} \, \frac{ \alpha_s \, \alpha_{em}  \, f_\rho^2 }{s}
\left( 4  \ln^2 \frac{Q_1 \, Q_2}{s} + 14 \, \ln \frac{Q_1 \, Q_2}{s} -4 \, \ln^2 Q_1/Q_2 +12 \right)\,. \nonumber
\eqa
Other amplitudes vanish at $t=t_{min}.$

These expressions should be compared with the corresponding 2 gluons exchange contributions
discussed in the previous sections.
The LL amplitude is almost constant around $t=t_{min},$ and given by (\ref{resultM00min}).

The TT amplitude (\ref{evaluateMij}) behaves as
\beq
\label{resultMTTmin}
\hspace{-1.1cm}{\cal M}^{gg}_{TT} \simeq  -i \, a \, \frac{\pi}{2} \, s \, \frac{N_c^2-1}{N_c^2} \, \alpha_s^2 \, \alpha_{em}  \, f_\rho^2 \,  \frac{|t-t_{min}|}{Q_1^3
  Q_2^3}\,,
\eq
where the constant $a=253.5$ is extracted from a numerical fit.

The Eqs. (\ref{MLLcollinear} - \ref{resultMTTmin}) confirm the well known fact that in the Regge limit 
the two gluon exchange dominates over the double quark exchange.
Indeed, the comparison of expressions (\ref{MLLcollinearS}) and (\ref{MTTcollinearS}) with formulas (\ref{resultM00min}) and (\ref{resultMTTmin}) shows that gluonic contributions are proportional to s
(in agreement with the usual counting rule $s^{\Sigma \sigma_i -N +1},$
where $N$ is the number of $t$ channel exchanged particles of spin $\sigma_i$). 
In the case of longitudinally polarized photons which does not vanish 
at $t_{min}$, and for the same photon virtualities  $Q_1^2=Q_2^2=Q^2$, let us consider the ratio
\beq
\label{ratioLL}
{\cal R}_{LL}= \frac{{\cal M}^{q \bar{q}}_{00}}{{\cal M}^{gg}_{00}} = \frac{32 \,(Q_u^2+Q_d^2)}{28 \, \zeta (3) -24 } \, \frac{Q^2}{s \, \alpha_s}\,.
\eq
For a typical value of $Q^2=1$ ${\rm GeV^2}$, as soon as $s$ ($\simeq s_{\gamma^*\gamma^*}$) is higher than    
4 ${\rm GeV^2},$ this ratio is bigger than unity, which at first sight seems to be always the case for ILC.
(\ref{resultM00min}) would thus completely dominates  with respect to (\ref{MLLcollinear}),
by several orders of magnitudes. In fact, $s_{\gamma^* \gamma^*}$ can reach such low value as 4 ${\rm GeV^2},$ because of the outgoing energy carried by the outgoing leptons and the strong peak of the Weizs\"acker-Williams fluxes at small $\gamma^*$ energies. 
We discuss this effect  in section \ref{eeresults} at the level of the $e^+e^-$ process, after performing the phase-space integration of the differential cross-section at $t_{min}$. It will be  shown that nevertheless the quark contribution is really negligible in almost all the ILC phase space.

In the case of the two gluon contribution with transverse virtual photons (\ref{resultMTTmin}) which vanishes at $t=t_{min}$, its dominance over the corresponding 
quark contribution  (\ref{MTTcollinear}) appears very rapidly when $|t-t_{min}|$ starts to 
increase, and persists in the whole essential region of the phase space (remember that (\ref{resultMTTmin}) is peaked at $t-t_{min}=k \, \,  0.01\rm GeV^2$ where $k$ is of order 1-10).
This dominance will also be discussed in more detail
in  section \ref{eeresults} at the level of the $e^+e^-$ process.




\section{Non-forward Born order cross-section for
 $e^+e^- \to e^+e^- \rho_L^0  \;\rho_L^0$ }
\label{crossee}

\setcounter{equation}{0}
\subsection{Kinematical cuts for the phase-space integration}
\label{cuts}

\begin{figure}
\centering{\includegraphics*[width=\wid]{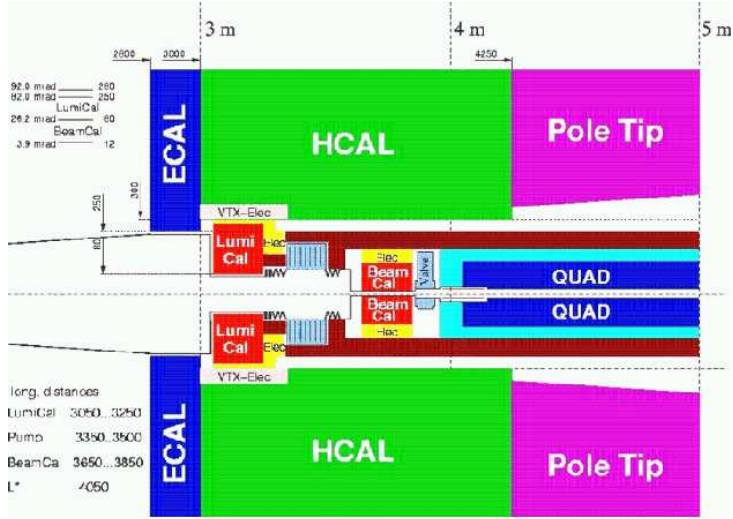}}
\caption{\small The LDC project, with the BeamCal forward electromagnetic calorimeter.}
\label{FigLDC}
\end{figure}

Our purpose is now to evaluate the cross-section of the process $e^+e^- \to e^+e^- \rho_L^0  \;\rho_L^0$ in the planned  experimental conditions 
of the International Linear Collider  project \cite{ilc}.
For the detector part, we chose to focus on the Large Detector Concept  \cite{LDC}, and use the potential of the very forward
region accessible through the electromagnetic calorimeter  BeamCal which may be installed
around the beampipe at 3.65 m from the interaction point. The LDC is illustrated in Fig.\ref{FigLDC}.

The cross-section which takes into account all the kinematical constraints, which are explained below, is given by
\bea
\label{sigma}
\frac{d\sigma^{e^+ e^- \to e^+ e^- \rho_L \rho_L}}{dt}\!=\!
\int^{Q_{1max}^2}_{Q_{1min}^2}dQ_{1}^2 \, \int^{Q_{2max}^2}_{Q_{2min}^2}dQ_{2}^2 \, \int^{y_{max}}_{\eps} \!\! dy_{1}\,  \int^{y_{max}}_{\frac{Q_1 Q_2}{s y_1}} \!\! dy_{2}\,\frac{d\sigma^{e^+ e^- \to e^+ e^- \rho_L \rho_L}}{dt \, dy_{1}\,dy_{2}\,dQ_{1}^2\,dQ_{2}^2}\, ,
\eea
with
$Q_{1min}=1$ GeV, $Q_{1max}=4$ GeV, $\eps=10^{-6}$ and $y_{max}=0.6$.

The cross-section (\ref{sigma}) can be evaluated combining the cross-section formulae (\ref{eesigma}), (\ref{crosssection}) and the results of section \ref{results} for the helicity scattering amplitudes.


The important feature of the formula  (\ref{sigma}) is that the dominant contribution
for the $e^+ \, e^- \, \to e^+ \, e^- \,  \rho_L^0  \, \rho_L^0$ process is strongly peaked at low $Q_i.$
The integration over $Q_i,$ $y_i$ is peaked in the low $y_i$ and $Q_i$ phase space
region due to the presence in (\ref{eesigma}) of $1/(y_i \, Q_i^2)$ factors coming from the Weizs\"acker-Williams fluxes, and thus amplifies this effect.
We show below that this dominant part of the phase space is accessible experimentaly using the BeamCal
calorimeter.

The integration domain in (\ref{sigma}) is fixed by the following considerations.
In the laboratory frame, which is also the center of mass system (cms) for a linear collider, the standard expression for the momentum fractions which respect to the incoming leptons and for the virtualities of the bremsstrahlung photons  are, respectively, given by
\bea
\label{yi}
y_i =  \frac{E-E_i' \cos^2(\theta_i/2)}{E} \quad {\rm and} \quad Q_i^2 =  4 E E_i' \sin^2(\theta_i/2) \, ,
\eea
where $E$  is the energy of the beam, while $E_i'$  and $\theta_i$ are respectively the energy and the scattering angle  of the out-going leptons. At ILC, the foreseen cm energy is $ \sqrt{s}=2 E=500$ GeV.
The experimental constraint coming from  the minimal detection angle $\theta_{min}$ around the beampipe is given by
 $\theta_{max}=\pi-\theta_{min}>\theta_i>\theta_{min}$ and leads to the following constraint on $y_i$
\bea
\label{yimin}
y_i > f(Q_i) =1-  \frac{Q_i^2}{s \tan^2(\theta_{min}/2)} \,,
\eea
where the constraint on the upper bound of $y_i$ coming from $\theta_{max}$ is completely negligible at this cm energy.\\
The condition on the energy of the scattered lepton $E_{max}>E_i'>E_{min}$ results in 
\bea
\label{yiE}
y_{i\, max} = 1-\frac{E_{min}}{E}>y_i >1-\frac{E_{max}}{E} .
\eea
Moreover we  impose that $s_{\gamma^*\gamma^*}=y_1 y_2 s>c \, Q_1 \, Q_2$ (where $c$ is an arbitrary constant of the order 1) which is required by the Regge kinematics for which the impact representation is valid. 
In subsection \ref{eeresults} we show that this constant $c$ can be adjusted to
choose bins of data for which  also in the case
of $e^+ e^-$ scattering the contribution with quark exchanges (discussed in 
sec.\ref{subdominant}) is completely negligible.

We arbitrarily choose $Q_i$ to be bigger than 1 GeV as it provides the  hard scale of the process which legitimates the use of perturbation theory. $Q_{i\, max}$ will be fixed to 4 GeV, since the various amplitudes
involved are completely negligible for higher values of virtualities $Q_i$ values (see section \ref{results}).
The constraints on $y_{i\,min}$ discussed so far 
are summarized by conditions 
\bea
\label{y12min}
\!\! y_{1\,min}= \max \left( f(Q_1),1-\frac{E_{max}}{E}\right) \ {\rm and} \
y_{2\,min}=\max \left( f(Q_2), 1- \frac{E_{max}}{E}, \frac{c \, Q_1 \, Q_2}{s\, y_1} \right)\,.
\eea
Further simplifications of conditions (\ref{y12min}) can be done by taking into 
account that the only condition on the maximal value of energy detection of the scattered leptons comes from kinematics, i.e. $E_{max}=E,$ and 
some specific features of the planned detector.

\begin{figure}
\epsfxsize=\wids{\centerline{\epsfbox{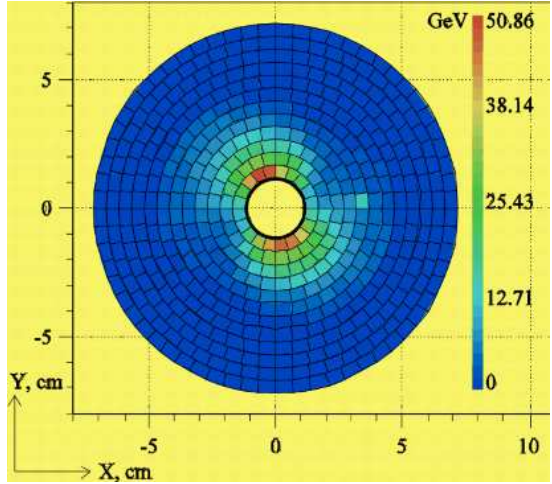}}}
\caption{\small Energy deposit in GeV in each cell of the BeamCal detector due to beamstrahlung.}
\label{Figdetector}
\end{figure}

The  BeamCal calorimeter in the very forward region allows in principle to detect particles down to 4 mrad.\footnote{Note that  in order to get access to any  inclusive \cite{royon} or diffractive high $s$ processes, one needs very small $\theta_{min}$. To reach  values of $\theta_{min}$ of a few mrads represents an important technological step
which was not feasible a few years ago.}  More precisely, it measures  an energy deposit for an angle between 4 mrad and 26 mrad. But this detector is also polluted by the photon beamstrahlung, specialy for very small angles (see Fig.\ref{Figdetector}). We assume
a non ambiguous identification for particles whose energies are bigger than 100 GeV. 
More precisely, the efficiency of detection of an electron
 depends on its energy and becomes less ambiguous when the energy increases. It
 is above 70 \% in the part of the phase space 
  which dominates the cross-section (small $y_i$, corresponding to $E'_i \simeq E_i$).
  A precise evaluation of this efficiency would require to set up a Monte Carlo simulation for the beamstrahlung 
  contribution, which is beyond the scope of this paper.
This sets the maximal value of $y_i$ to $y_{i\,max}=1-\frac{E_{min}}{E}=0.6$ with $E_{min}=100$  GeV and $ E=250$ GeV.

\begin{figure}
\begin{picture}(800,155)
\put(0,0){\epsfxsize=\widm{\centerline{\epsfbox{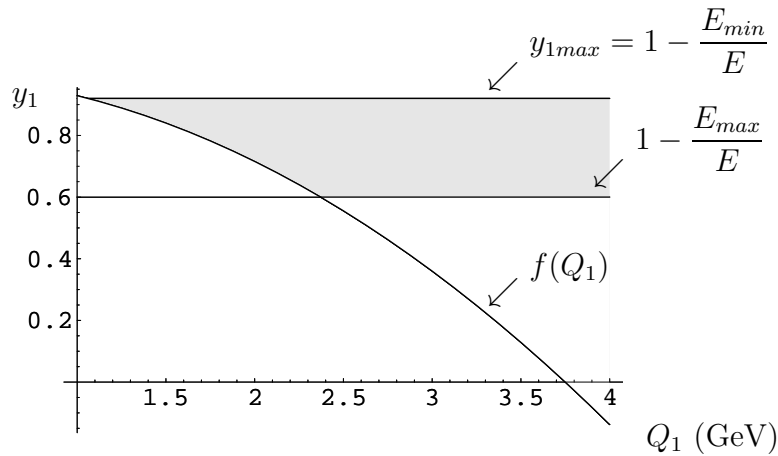}}}}
\put(290,55){$\swarrow$ \raisebox{0.8 \totalheight}{$f(Q_1)$}}
\put(290,137){$\swarrow$ \raisebox{0.5 \totalheight}{$\displaystyle y_{1max}=1-\frac{E_{min}}{E}$}}
\put(330,100){$\swarrow$ \raisebox{0.5 \totalheight}{$\displaystyle 1-\frac{E_{max}}{E}$}}
\put(350,0){$Q_1$ (GeV)}
\put(110,130){$y_1$}
\end{picture}
\caption{\small $y_1$ integration domain for $\theta_{min}=10$ mrad, $E_{min}=20 \, \rm GeV$ and $E_{max}=100 \,  \rm GeV$.}
\label{Figf12}
\end{figure}

Such a big value of  $E_{min}$ can be considered as surprisingly high and could 
lead to a strong 
reduction of the allowed phase space. In principle one could enlarge the phase space by taking into account particles whose energies $E_i'$ are between 100 GeV and 20 GeV with angles $\theta_i$ bigger than 10 mrad (see Fig.\ref{Figdetector}), 
but the contribution of this domain is negligible (see Fig.\ref{Figf12}) since the lower bound of $y_i$ (see  Eq.(\ref{y12min}))
prevents us to reach the small values of $y_i$ and $Q_i$ which give the dominant contribution to the cross-section. We safely neglect the contribution of 
this region of phase space and assume in the following $E_{min}=100$ GeV and $\theta_{min}=4$ mrad. 

Thus, with $\theta_{min}=4$ mrad and $ \sqrt{s}=500$ GeV, we  
 have $s \tan^2(\theta_{min}/2)=1\, \rm GeV^2$, which means that $f(Q) \leq 0$ for $Q^2 \, \geq 1 {\rm GeV^2}.$ The relations (\ref{y12min}), with $E_{max}=E,$ reduce to only one condition
 $y_{2min}=\frac{Q_1 Q_2}{s y_1}$. This has to be supplemented numerically with the condition $y_{1min}=\eps$, where $\eps$ is a numerical cut-off: although, because of the Regge limit condition, we have 
$y_1> \frac{Q_1 Q_2}{s y_2}\geq\frac{Q_{1min} Q_{2min}}{s y_{2max}}=6.6 10^{-6}$ which thus provides a natural lower cut-off for $y_1,$ nevertheless we choose $\eps=10^{-6}$ so that it is smaller than the smallest reachable value of $y_{1}$ but still non zero.
This cut-off has no practical effect, except for
avoiding numerical instabilities in the integration code.

The above discussion justifies the various cuts in formula (\ref{sigma}).

\subsection{Background in the detector}
\label{background}

BeamCal is an electromagnetic calorimeter which cannot distinguish charges of particles. Thus,
it is important 
to check that the cross-sections of any other processes which could lead 
to 
final states which can be misidentified with the final state of the process $e^+e^- \to e^+e^- \rho_L^0  \;\rho_L^0$
 are suppressed.
Indeed, the final state of 
the process  $e^+e^- \to \gamma \gamma \rho_L^0  \;\rho_L^0$, 
with photons of the same energy deposit in detector as outgoing leptons, 
cannot  be distinguished 
with the final state of $e^+ e^- \to e^+ e^- \rho^0_L \rho^0_L$.

\begin{figure}
\begin{picture}(800,170)
\put(-120,23){\epsfxsize=\widss{\centerline{\epsfbox{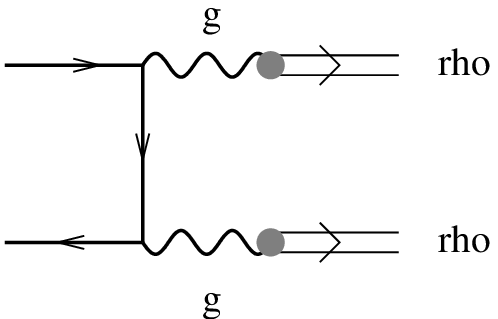}}}}
\put(120,0){\epsfxsize=\widss{\centerline{\epsfbox{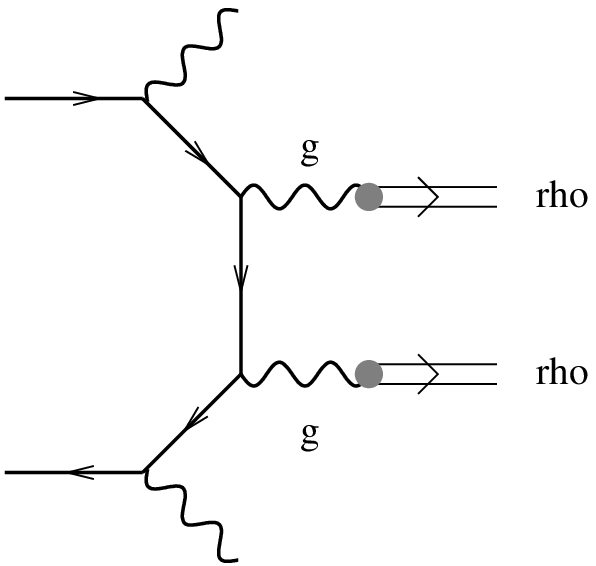}}}}
\put(140,0){$(a)$}
\put(380,0){$(b)$}
\end{picture}
\caption{\small Example of Born order diagrams for the process $e^+e^- \to  \rho_L^0  \;\rho_L^0$ (a) and for the $e^+e^- \to \gamma \,\gamma \,\rho_L^0  \;\rho_L^0$ process (b).}
\label{Figeerhorho}
\end{figure}

We shall argue that the process $e^+e^- \to \gamma \gamma \rho_L^0  \;\rho_L^0$
leads to a cross-section which is negligible at ILC.
Let us first start with the process  $e^+ e^- \to \rho^0_L \rho^0_L$
illustrated in Fig.\ref{Figeerhorho}(a), studied in Ref.\cite{davier,braaten}. 


\psfrag{rho}[cc][cc]{$\rho_L^0$}
\psfrag{g}[cc][cc]{$ $}
Its differential cross-section behaves typically like
\beq
\label{eerhorho}
\frac{d \sigma}{d t} \varpropto \frac{\alpha_{em}^4 f_{\rho}^4}{s^2 m_{\rho}^4}\,,
\eq
with the virtualities of the photons propagators equal to $m_\rho^2.$ More accurate expressions can be found in \cite{braaten},  if one identifies $g_{\,V\gamma}=f_\rho \,  m_\rho$.

Now, when considering the competitor process $e^+e^- \to \gamma \gamma \rho_L^0  \;\rho_L^0,$
that is adding two additional bremsstrahlung photon as in Fig.\ref{Figeerhorho}(b),
we get
\beq
\label{ratio}
\frac{d\sigma^{\,e^+ e^- \to \gamma \gamma \rho_L \rho_L}}{dt \, dy_{1}\,dy_{2}\,dQ_{1}^2\,dQ_{2}^2} / \frac{d\sigma^{\,e^+ e^- \to e^+ e^- \rho_L \rho_L}}{dt \, dy_{1}\,dy_{2}\,dQ_{1}^2\,dQ_{2}^2}  \simeq \frac{\alpha_{em}^2 Q_1^4 Q_2^4}{\alpha_s^4 s^2 m_\rho^4}
\eq
which is suppressed at ILC energies, and would be of comparable order of
magnitude only for colliders with cm energy of the order of a few GeV.

\subsection{Results for cross-section}
\label{eeresults}

\begin{figure}[h]
\begin{picture}(10,240)
\put(0,10){\epsfxsize=11cm{\centerline{\epsfbox{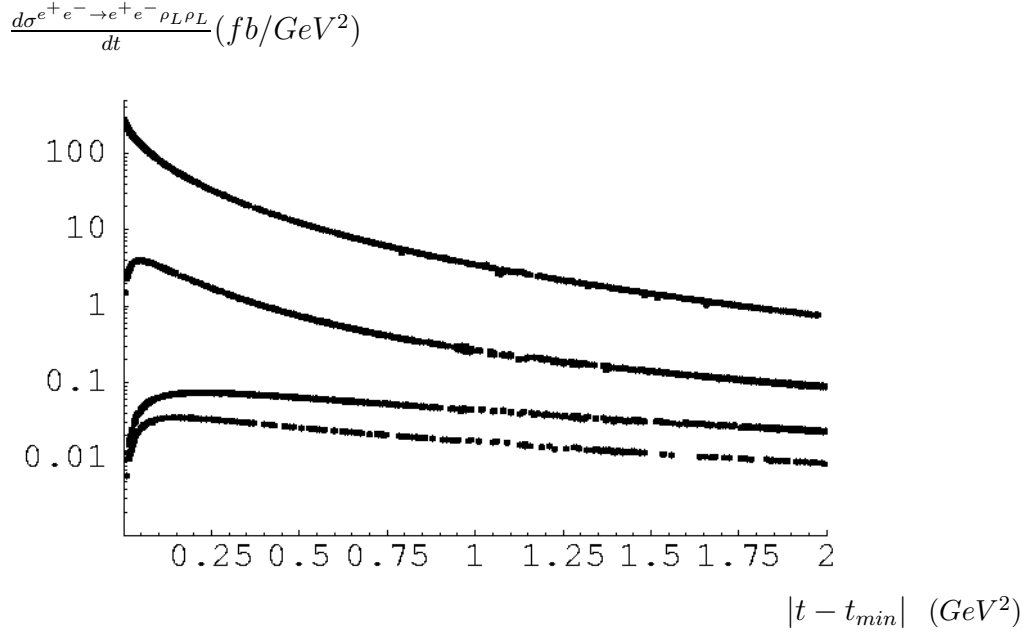}}}}
\put(360,0){ $|t-t_{min}| $ \  \small $ (GeV^2)$}
\put(70,220){$\frac{d\sigma^{e^+ e^- \to e^+ e^- \rho_L \rho_L}}{dt} (fb/GeV^2)$}
\end{picture}
\caption{\small Cross-sections for $e^+e^- \to e^+e^- \rho_L^0  \;\rho_L^0$ process. Starting from above, we display the
 cross-sections  corresponding to the  $\gamma^*_L \gamma^*_L$ mode,  to the $\gamma^*_L\gamma^*_T$ modes,   to the 
$\gamma^*_T \gamma^*_{T'}$ modes with  different $T \neq T'$ and finally to the $\gamma^*_T \gamma^*_{T'}$ modes with  the same $T=T'$.}
\label{FigLogcurves}
\end{figure}

\begin{figure}[h]
\begin{picture}(10,240)
\put(0,10){\epsfxsize=11cm{\centerline{\epsfbox{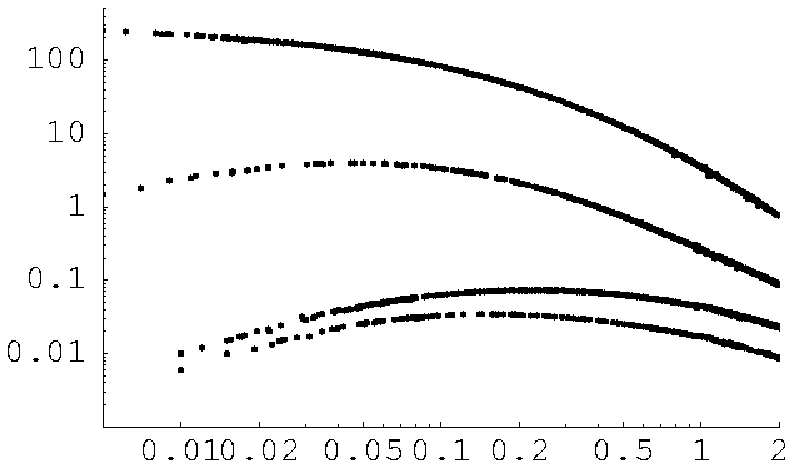}}}}
\put(360,0){ $|t-t_{min}|$ \  \small $ (GeV^2)$}
\put(70,220){$\frac{d\sigma^{e^+ e^- \to e^+ e^- \rho_L \rho_L}}{dt} (fb/GeV^2)$}
\end{picture}
\caption{\small Cross-sections for $e^+e^- \to e^+e^- \rho_L^0  \;\rho_L^0$ process, in log-log scale.
Starting from above, we display the
 cross-sections  corresponding to the  $\gamma^*_L \gamma^*_L$ mode,  to the $\gamma^*_L\gamma^*_T$ modes,   to the 
$\gamma^*_T \gamma^*_{T'}$ modes with  different $T \neq T'$ and finally to the $\gamma^*_T \gamma^*_{T'}$ modes with  the same $T=T'$.
}
\label{FigLogLogcurves}
\end{figure}

We now display in Fig.\ref{FigLogcurves} the cross-sections $\frac{d\sigma^{e^+ e^- \to e^+ e^- \rho_L \rho_L}}{dt} $ as a function of $t$ for the different polarizations, which are plotted after integrating the differential cross-section in  (\ref{sigma})  over the phase space considered previously.
We made the following assumptions: we choose the QCD coupling constant to be $\alpha_{s}(\sqrt{Q_1 Q_2})$ running at three loops, the parameter $c=1$ which enters in the Regge limit condition and the cm
 energy $ \sqrt{s}=500$ GeV.
Fig.\ref{FigLogcurves} shows for $e^+e^-$ scattering the same differential cross-sections 
related to different photon helicities as Fig.\ref{FigQGeV}. We see that the shapes 
of corresponding curves are similar although they lead to quite different 
values of cross-sections. The cross-sections corresponding to photons with 
at least one transverse polarization  vanish as in 
the $\gamma^*\gamma^*$ (cf subsection \ref{results}) case at $t=t_{min}.$ Similarly, each of them has a maximum in the very small $t$ 
region. These maxima are shown more accurately on the log-log plot in Fig.\ref{FigLogLogcurves}.

At this point one technical remark is in order.
By looking into the upper 
plot in Fig.\ref{FigLogLogcurves} related to the ${\cal M}_{00}$ amplitude, one sees that the points 
corresponding to nonzero $|t-t_{min}|$ approach smoothly the point on the axis $|t-t_{min}|=0$. 
This point  $|t-t_{min}|=0$ is of special interest because it gives the maximum of the total cross-section (since the transverse polarization case vanishes at $t_{min}$) and then practically dictates the trend of the total cross-section which is strongly peaked in the forward direction (for the longitudinal case) and strongly decreases with $t$ (for all polarizations), as shown already at the level of the $\gamma^* \gamma^*$ cross-sections in subsection \ref{results}.
Due to numerical instabilities, the differential cross-section at $|t-t_{min}|=0$  must be evaluated in a different 
way than those for $|t-t_{min}| \neq 0$, i.e. by the use of expression  (\ref{resultM00min}) in 
which the integration 
over $z_i$ was already done in the analytic way. Since Eq.(\ref{resultM00min}) 
involves several polylogarithmic functions its structure of cuts  
is quite inconvenient for further numerical integration over variables $y_i$ and $Q_i$.
In order to overcome this technical problem it is useful to rewrite (\ref{resultM00min})
by the use of Euler identity \cite{lewin} in the form
\bea
\label{newM00tmin}
&&\hspace{-1.1cm}{\cal M}_{00}= -i s \, \frac{N_c^2-1}{N_c^2} \, \alpha_s^2 \, \alpha_{em}  \, f_\rho^2 \, \frac{9 \pi^2}{2} \, \frac{1}{Q_1^2
  Q_2^2}  \left[6 \, \left(R + \frac{1}{R}\right) \ln^2 R  \right.\\
&&\hspace{-1cm}\left. + \,12\,
\left(R-\frac{1}{R}\right) \ln R \,+\, 12 \,  \left(R + \frac{1}{R}\right)\,+\,
\left(3 \,R^2 +\, 2\,+\frac{3}{R^2}\right) \left(\, \left(\frac{\pi^2}{6} -{\rm Li_2} \,(1-R)\, \right)   \,\ln R
  \nonumber \right. \right.\\
&&\hspace{-1cm}\left. \left.-\ln \,(R+1)\,\ln^2 R \,-\,2 \,{\rm Li_2}\,(-R)\,
    \ln R \,+\,{\rm Li_2}\,(R)\, \ln R \,+\,2\,{\rm Li_3}\,(-R)\,-\,2\, {\rm Li_3}\,(R)\right)\right]\,.\nonumber
\eea
since now the imaginary terms only come from ${\rm Li_2}\,(R)$ and ${\rm Li_3}\,(R)$ along their cuts, which cancels among each other analytically. Therefore, one can safely use their real part in a numerical fortran code as defined in standard packages.

The ILC collider is expected to run at a cm nominal energy of 500 GeV, though it might be extended in order to cover a range between 200 GeV and 1 TeV. 
Because of this possibility,
we below discuss how the change of the energy in cms influences our 
predictions for the cross-sections measured in the same BeamCal detector.
Furthermore, we discuss the effects of our various assumptions on the
cross-section $\frac{d\sigma^{e^+ e^- \to e^+ e^- \rho_L \rho_L}}{dt} $ at the point $t_{min}$, and 
consequently on the behaviour of the total cross-section.\\
\begin{figure}[h]
\begin{picture}(10,210)
\put(0,10){\epsfxsize=10cm{\centerline{\epsfbox{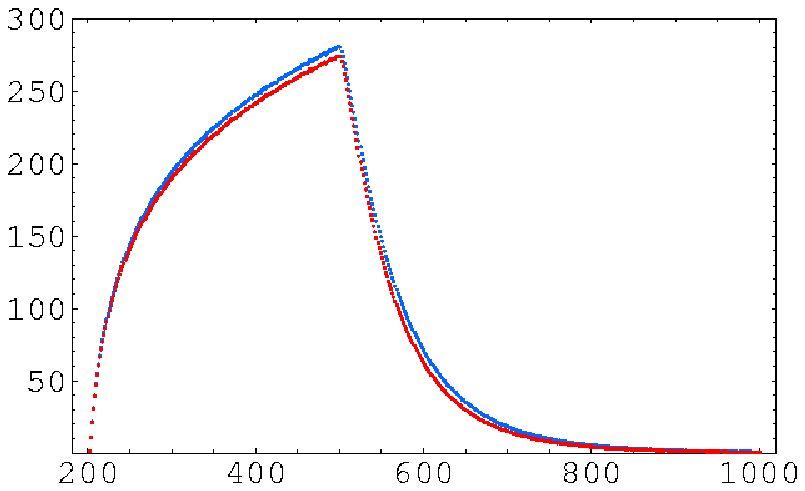}}}}
\put(370,0){ $\sqrt{s} $ \  \small $ (GeV)$}
\put(70,200){$\frac{d\sigma^{tmin}}{dt} (fb/GeV^2)$}
\end{picture}
\caption{\small Cross-sections for $e^+e^- \to e^+e^- \rho_L^0  \;\rho_L^0$ at $t=t_{min}$ for different $\alpha_s$ : the blue and  red curves for $\alpha_s$ running respectively at one and three loops, with $c=1.$   }
\label{Figborntmin}
\end{figure}

Fig.\ref{Figborntmin} shows the cross-section at $t_{min}$ as a function of the cm energy $\sqrt{s}$ for different choices of strong coupling constant $\alpha_s$.
 To see the sensitivity of our predictions to these choices, we plot the cross-section at $t_{min}$ in two cases: the blue curve corresponds to  $\alpha_{s}(\sqrt{Q_1 Q_2})$ running at one loop and the red one to $\alpha_{s}(\sqrt{Q_1 Q_2})$ running at three loops. The curves in Fig.\ref{Figborntmin} are very close to each other, which leads to a small uncertainty on the total cross-section as we will see in the following.

The shapes of plots in Fig.\ref{Figborntmin} distinguish two different domains:  if the planned cm energy range $\sqrt{s}$ is lower than 500 GeV, the function $f(Q_i)$ (cf. equation (\ref{yimin})) appearing as a constraint on the minimum value of $y_i$ in the phase space integration domain does not play any role at $\theta_{min}=4$ mrad. Thus the cross-section increases with $\sqrt{s}$ between 200 and 500 GeV.
Because of the condition we assumed on the minimal value of the energies 
of the scattered leptons in the section \ref{cuts}, the $y_i$ integration domain 
becomes very narrow (cf. equation (\ref{yiE})) when $\sqrt{s}$ goes to 200 GeV and leads to a strong 
decreasing of the cross-section at this cm energy.
Note that if $\sqrt{s}$ becomes bigger than 500 GeV, $f(Q_i)$ will cut the small $y_i$ region (which contribute mainly because of the Weizs\"acker-Williams photons fluxes) when $\sqrt{s}$ increases. Thus the cross-section falls down between 500 GeV and 1 TeV. This is due to the limitation caused by the minimal detection angle offered by the BeamCal calorimeter, which is thus optimal for our process when $\sqrt{s} = 500$ GeV.
This effect on $f(Q_i)$ could be compensated if one could increase the value of $Q_i$ but this would be completely suppressed because of the strong decreasing of the amplitude with $Q_i$.
The above discussion leads also to the conclusion, that although the Born order cross-sections do not depend on $s,$ the triggering effects introduce an  $s$-dependence of the measured cross-sections.

\begin{figure}[h]
\begin{picture}(10,200)
\put(0,10){\epsfxsize=10cm{\centerline{\epsfbox{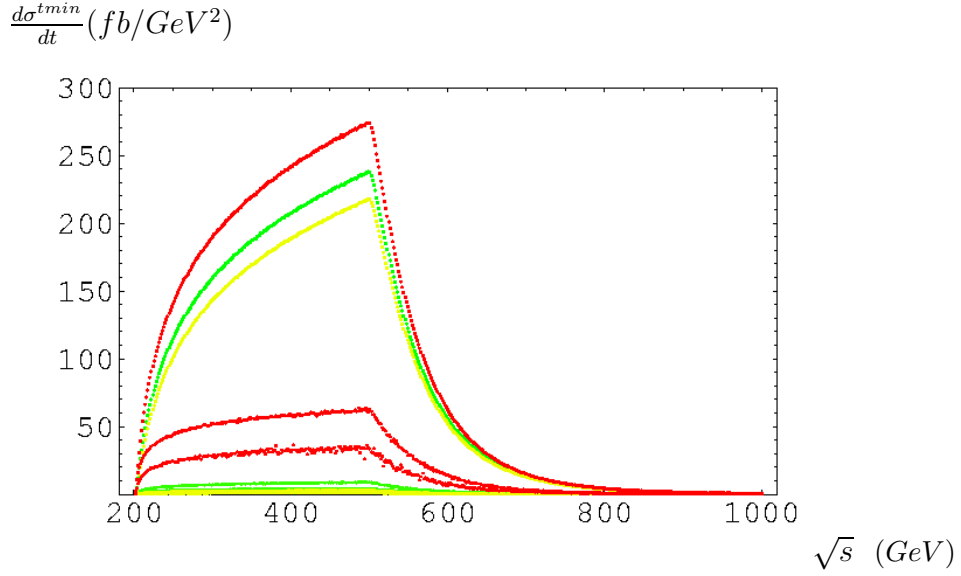}}}}
\put(370,0){ $\sqrt{s} $ \  \small $ (GeV)$}
\put(70,200){$\frac{d\sigma^{tmin}}{dt} (fb/GeV^2)$}
\end{picture}
\caption{\small Cross-sections for $e^+e^- \to e^+e^- \rho_L^0  \;\rho_L^0$ at $t=t_{min}$ for different values of the parameter $c$: the red [black] curves correspond to  $c=1$, the green [dark grey] curves to $c=2$ and and the yellow [light grey] curves to $c=3$. For each value of c, by decreasing order the curves correspond to gluon-exchange, quark-exchange with longitudinal virtual photons and quark-exchange with transverse virtual photons. }
\label{Figceffects}
\end{figure}

Fig.\ref{Figceffects} shows the cross-section at $t_{min}$ for different values of the parameter c which enters in the Regge limit condition $s_{\gamma^*\gamma^*}=y_1 \, y_2 \, s>c \, Q_1 \, Q_2$. The value of the parameter c controls the dominance of gluonic contributions to the scattering amplitude: the increase of c should lead to suppression of quark exchanges. To see that  we  display the quark contribution in the same bins: we use the usual phase-space for the process $e^+e^- \to e^+e^- \rho_L^0  \;\rho_L^0$ (cf. Eq.(\ref{eesigma})) with the expressions of the amplitudes (\ref{MLLcollinear}) and (\ref{MTTcollinear}), and perform their numerical integration on $y_i$, $Q_i$ with the same cuts as in the two gluon exchange process (cf. section \ref{cuts}). For each value of c we plot the three curves corresponding to the  two gluon exchange process and  the quark exchange processes with longitudinal and transverse virtual photons.

A technical remark is in order when performing this integration numerically. The equation (\ref{MTTcollinear}) is not divergent when $Q_i^2 \to s $ because  this limit is only valid if  $s (1-Q_1^2/s)(1-Q_2^2/s)$ is finite and positive since this term corresponds in our notation to the cm energy of the virtual photons. In order to avoid numerical instabilities we add the condition $ y_1 y_2 s> Q_1^2 ,\, Q_2^2$ to the Regge limit condition. We can check that this supplementary constraint does not change our results for the other contributions, namely for the  two gluon exchange and the quark exchange with longitudinal virtual photons processes. 

As expected, the quark contribution is suppressed when increasing $c$ and becomes completely negligible as soon as $c$ exceeds 2.
All above discussion concerned the case  $t=t_{min}$ which determines the general trend of the cross-section in the non forward case. Because of that we hope that above conclusions are also valid at the level of the integrated over $t$ cross-section. 
Thus, we omit bellow the quark exchanges.

We finally obtain the following results for the total cross-section integrated over $t$. We shall show three different predictions which differ by the choice of the definition of the coupling constant and by the choice of the value of the parameter c controlling the gluon dominance. First we choose $\alpha_{s}(\sqrt{Q_1 Q_2})$ running at three loops, the constant $c=1$, the cm energy $\sqrt{s}=500$ GeV and we obtain (up to numerical uncertainties):
\bea
\sigma^{LL}&=&32.4 \,  fb \\ \nonumber
\sigma^{LT}&=&1.5 \,   fb \\ \nonumber
\sigma^{TT}&=&0.2  \,  fb
\label{sigmatotalc1}
\eea
and then
\bea
\sigma^{Total}=34.1  \,  fb.
\eea
With a nominal integrated luminosity of $125  \, {\rm fb}^{-1},$ this will yield $4.26\, 10^3$ events per year.

 Secondly, with the choice of  $\alpha_{s}(\sqrt{Q_1 Q_2})$ running at one loop, the constant $c=1$ and the cm energy $ \sqrt{s}=500$GeV, we obtain:
\bea
\sigma^{LL}&=&33.9  \,  fb \\ \nonumber
\sigma^{LT}&=&1.5 \,   fb \\ \nonumber
\sigma^{TT}&=&0.2  \,  fb
\eea
which leads to
\bea
\sigma^{Total}=35.6  \,  fb.
\eea
As expected, we see that the transition from three to one loop changes very little the total cross-section.
This result will yield $4.45\, 10^3$ events per year with a nominal integrated luminosity of $125  \, {\rm fb}^{-1}$.

 In the third choice, we choose $\alpha_{s}(\sqrt{Q_1 Q_2})$ running at three loops, the same cm energy $ \sqrt{s}=500$GeV and the constant $c=2$ (for which as previously discussed quark exchanges are completely negligible) and we get: 
\bea
\sigma^{LL}&=&28.1 \,  fb \\ \nonumber
\sigma^{LT}&=&1.3 \,   fb \\ \nonumber
\sigma^{TT}&=&0.2  \,  fb
\eea
which imply 
\bea
\label{sigmatotalc2}
\sigma^{Total}=29.6  \,  fb.
\eea
This result  will yield $3.7\, 10^3$ events per year with a nominal integrated luminosity of $125 \, {\rm fb}^{-1}$.

Finally, we also consider the same assumptions as the previous ones except for the value of the constant $c$ which is now set to $c=10$ in order to consider a more drastic Regge limit condition and we obtain:

\bea
\sigma^{LL}&=&19.3 \,  fb \\ \nonumber
\sigma^{LT}&=&0.9 \,   fb \\ \nonumber
\sigma^{TT}&=&0.11  \,  fb
\eea
which leads to 
\bea
\label{sigmatotalc10}
\sigma^{Total}=20.3  \,  fb.
\eea
This result  will yield $2.5\, 10^3$ events per year with a nominal integrated luminosity of $125 \, {\rm fb}^{-1}$. Thus, this shows that the precise way one implements the restriction of the kinematical phase space to the domain of applicability of the impact representation does not dramatically change the number of events.

All the prediction above were obtained using the asymptotical DA.
In order to see the sensitivity of this assumption on our results, we do also the calculation
using the DA (\ref{asymptDA}) within different models.
The choice of the DA of Ref.\cite{polyakov} with $a_2 = -0.1$ and $a_4=0$
gives $4.2 \, 10^3$ events per year, while the choice of the DA of Ref.\cite{BM} with $a_2 = 0.05$ and $a_4=0$
gives $4.3 \, 10^3$ events per year.

 In summary of this part we see that our predictions are quite stable when changing the main parameters characterizing the theoretical uncertainties of our approach. 
 
The obvious question which appears now is how our predictions summarized by
 Eqs. (\ref{sigmatotalc1}-\ref{sigmatotalc2}) will change by the inclusion of the BFKL resummation effects. Generaly BFKL evolution increases strongly values of cross-sections, which means that
usually the results obtained at Born approximation can be considered as a lower limit of cross-sections for $\rho$-mesons pairs production with complete BFKL evolution taken into account. Although the complete analysis of BFKL evolution  for our process is beyond the scope of the present paper, we would like to finish this section with  a few remarks on possible effects caused by the BFKL evolution.

We consider below only the point $t=t_{min}$ and we restrict ourselves to the leading order BFKL evolution\footnote{Note that related studies with a hard scale provided by $t$ and not by $Q^2$ were performed in \cite{BFKLlarget}}.  Of course such an estimate should be taken with great caution
since it is well known that LO BFKL overestimates the magnitude of corrections.
\begin{figure}[h]
\begin{picture}(10,220)
\put(0,10){\epsfxsize=10cm{\centerline{\epsfbox{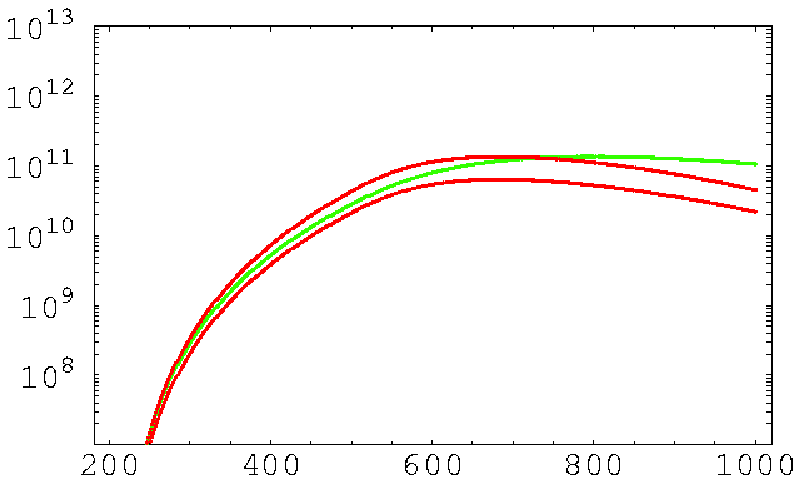}}}}
\put(370,0){ $\sqrt{s} $ \  \small $ (GeV)$}
\put(70,200){$\frac{d\sigma^{tmin}}{dt} (fb/GeV^2)$}
\end{picture}
\caption{\small Cross-sections for $e^+e^- \to e^+e^- \rho_L^0  \;\rho_L^0$ with LO BFKL evolution at $t=t_{min}$ for different $\alpha_s$ : the upper and lower  red curves for  $\alpha_s$ running respectively at one and three loops  and the green one (the middle curve) for $\alpha_s = 0.46$.  }
\label{Figbfkltmin}
\end{figure}

In the Fig.\ref{Figbfkltmin} we show the corresponding cross-section at $t_{min}$ 
 as a function of $\sqrt{s}$, for different choices of  $\alpha_{s}$: we considered  $\alpha_s$ running at one and three loops (red curves) as in the previous discussion for the two gluon exchange and we also used a fixed value of $\alpha_s$ (green curve) corresponding to the three loops running coupling constant at a typical virtuality $Q=1.1$ GeV. We have used  the expression of the BFKL amplitude  \cite{epsw} for the forward case in the saddle point approximation, namely
\beq
A(s,t=t_{min},Q_1,Q_2) \sim i s \, \pi^5\sqrt{\pi} \, \frac{9 (N_c^2-1)}{4 N_c^2}
\frac{\alpha_s^2 \alpha_{em} f_\rho^2}{Q_1^2 Q_2^2}
\frac{e^{4\ln 2 \; \as Y}}{\sqrt{14 \as \zeta(3) Y}}
\exp\left(-\frac{\ln^2 R}{14 \as \zeta(3) Y}\right)\,,
\label{AsaddleR}
 \eq
with the rapidity $Y=\ln(\frac{c' \, s \, y_1 \, y_2}{Q_1 Q_2})$, $ \as=\frac{N_c}{\pi} \alpha_{s}(\sqrt{Q_1 Q_2}) $ and $R=\frac{Q_1}{Q_2}$. The plots in Fig.\ref{Figbfkltmin} are obtained by assuming that the constant $c'$ in Eq.(\ref{AsaddleR}), which at LO is arbitrary and of order 1, is chosen to be 1. The factor $\exp(4\ln 2 \; \as Y)$ 
 explains the enhancement of the sensitivity to the choice of $\alpha_s$ compared to the one in the Born two gluon exchange case, since $4\ln 2 \; Y$  takes big values for ILC rapidities $Y.$ For the same reasons as discussed  earlier in this section, the function $f(Q_i)$ does not appear for $\sqrt{s}$ lower than 500 GeV; the LO BFKL cross-section then grows exponentially with $s$ in this domain. The effect of $f(Q)$ starting from 500 GeV gives an inflexion point of the curves  and a maximum beyond 500 GeV; then the curves decrease until 1TeV.
 \begin{figure}[h]
\begin{picture}(10,220)
\put(0,10){\epsfxsize=10cm{\centerline{\epsfbox{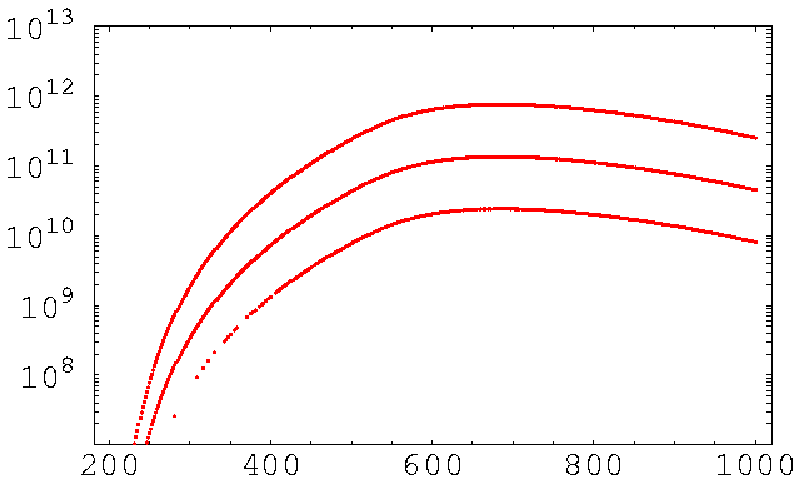}}}}
\put(370,0){ $\sqrt{s} $ \  \small $ (GeV)$}
\put(70,200){$\frac{d\sigma^{tmin}}{dt} (fb/GeV^2)$}
\end{picture}
\caption{\small LO BFKL cross-section for $e^+e^- \to e^+e^- \rho_L^0  \;\rho_L^0$ at $t=t_{min}$ for different values of the parameter $c'$: by decreasing order, the curves correspond to  $c'=2$, $c'=1$ and  $c'=0.5\,.$ $c$ is fixed to be equal to 1.}
\label{Figc'effects}
\end{figure}

The effect of varying the parameter $c'$ in the BFKL prediction is illustrated in Fig.\ref{Figc'effects}.
As expected, it has a strong effect in the order of magnitude of the differential cross-section, since the rapidity is very high and thus leads to a large value of the factor $\exp(4\ln 2 \; \as Y)$, which is highly sensitive to the precise definition of the rapidity.

Comparing the order of magnitude of Born cross-section (Fig.\ref{Figborntmin} and \ref{Figceffects}) with cross-sections provided by the LO BFKL evolution (Fig.\ref{Figbfkltmin} and Fig.\ref{Figc'effects}), one could be astonished by the fact that they differ by several orders of magnitudes.
From previous studies at the level of $\gamma^*\gamma^*$ \cite{epsw},\cite{ivanov}, the NLO contribution is known to be between LO and Born order cross-section. 
 Thus, at the level of the $e^+e^-$ process, such a large magnitude for the LO BFKL cross-section will be suppressed at NLO, leading to a more realistic estimate.

The above discussion about BFKL enhancement was restricted to the forward case $t=t_{min}$.   
In the non-forward case, the phase space region with small $t$ values dominates  the cross-sections.  The obtained hierarchy between cross sections in Born approximation for
different photon polarizations will presumably still be valid when including BFKL evolution at any order of resummation (LO, NLO, etc...).
Indeed the argument given in section \ref{results} for Born order and on which this hierarchy 
is based, only relies on the $s$-channel helicity conservation. Technically, it is based on the impact representation which is valid beyond Born and/or LO approximation.

The comparison of Figs.\ref{Figborntmin}-\ref{Figceffects} with  Figs.\ref{Figbfkltmin}-\ref{Figc'effects}  leads to the conclusions that the BFKL evolution changes the shape of the cross-section:
 when increasing $\sqrt{s}$
 from 500 GeV to 1 TeV, the two gluon exchange cross-section will fall down, while with  the BFKL
resummation effects, the cross-section should more or less stay stable, with a high number of events to be still observed for these cm energies.


\section{Conclusion}

The present study should be considered as a continuation of our
previous investigations \cite{conf},\cite{psw},\cite{epsw} for the production of two
$\rho^0_L$-mesons in the scattering of two longitudinally
polarized virtual photons.  The diffractive production of a meson
pair is one of the gold plated processes which permit clean
studies of the BFKL dynamics at ILC. Our main motivation in
the present work was to estimate, in the Born approximation,  the
cross-section for production of $\rho^0_L$-meson pairs in the
$e^+e^-$ collisions occurring in the kinematical conditions of
future ILC. For this aim, we first calculated contributions,
missing up to now,
 which involve the helicity amplitudes with transversally polarized virtual photons. This was done in a mostly analytic way, by the use of  techniques developped in Ref.\cite{psw}. Having done, we calculated the cross-section for the
 electroproduction of $\rho^0_L$-meson pairs which takes into account  kinematical cuts imposed by the LDC design project for the  BeamCal detector.  By assuming a nominal value of the integrated luminocity, we predict (in
the numerical analysis of cross-sections) a production of at least
4 $10^3$ 
meson pairs per year, a  value which is sufficiently large to ensure a reliable data analysis.

We discussed a possible background process in the BeamCal detector
which can identify in a misleading way an outgoing lepton with a
photon. We predict that the cross-section for such a background
process is negligibly small at ILC energies.

Finally we discussed theoretical uncertainties of obtained
estimates. There are two main sources of them. The first one is
related to the assumptions we have made to characterize the Regge limit and the particular role played by the parameter c; we also observe a sensitivity of our results on the choice of the running coupling constant.

The second source of theoretical uncertainties of our estimates
is related to taking into account effects of the BFKL evolution.
Generally, an inclusion of the BFKL evolution increases
significantly the cross-section, as one sees from comparison of Fig.\ref{Figborntmin} with Fig.\ref{Figbfkltmin} obtained within
 the LO BFKL approach. On the other
hand it is known that this increase of predictions for 
cross-sections is smaller if the BFKL evolution is considered at the next-to-leading order. Because of that we can safely say that our
predictions should be considered as a lower limit of predictions
which are obtained  by taking into account BFKL effects at NLO. We
hope to consider this issue in our future publications.

In principle, the same techniques can be applied for the description of processes involving other final states,
both with positive charge parity exchange in $t-$channel,  e.g. $J/\Psi$ pairs, as well as 
negative charge parity, e.g. $\gamma^* \gamma^* \to \eta_c \eta_c$ \cite{ewerz}.

\vskip.8cm
\noindent
{\large \bf Acknowledgements:}
\\

 We thank  G.~Korchemsky, S.~Munier, B.~Pire, R.~Poeschl and F.~Richard
for many
discussions and comments. This  work  is  supported  by
the Polish Grant 1 P03B 028 28,
the
French-Polish scientific agreement Polonium.
L.Sz. is a Visiting Fellow of
the Fonds National pour la Recherche Scientifique (Belgium).

\section{Appendix}

\setcounter{equation}{0}


In this appendix we collect  the analytical expressions of the coefficients  $a (\rb;Q_1,Q_2;z_1,z_2)$, $b(\rb;Q_1,Q_2;z_1,z_2)$ and $ f (\rb;Q_1,Q_2;z_1,z_2) ,$ as well as  all the generic integrals which appear
in the computation of the Born amplitude in section \ref{cross}. The coefficients $a$, $b$ and $f$ can be expressed as combinations of several generic 
integrals
\bea
\label{a_old}
a (\rb;Q_1,Q_2;z_1,z_2) &=& \frac{1}{2} \left[ I_{3\mu_1\mu_2}(z_1,\,z_2) + I_{3\mu_1\mu_2}(\zb_1,\,\zb_2) - I_{3\mu_1\mu_2}(\zb_1,\,z_2) -  I_{3\mu_1\mu_2}(z_1,\,\zb_2) \right] \nonumber
\\
&&\hspace{-1.5cm}- \frac{r^2}{2} \left[ I_{4\mu_1\mu_2}(z_1,\,z_2) - I_{4\mu_1\mu_2}(\zb_1,\,z_2) \right] \nonumber
\\
&&\hspace{-1.5cm}- \frac{1}{r^2}  \left[ J^c_{\mu_1 \mu_2}(z_1,\,\zb_2) + J^c_{\mu_1 \mu_2}(\zb_1,\,z_2) - J^c_{\mu_1 \mu_2}(\zb_1,\,\zb_2) - J^c_{\mu_1 \mu_2}(z_1,\,z_2) \right]\,,
\eea
\bea
\label{b_old}
b (\rb;Q_1,Q_2;z_1,z_2) &=& \left[\frac{z_1 z_2}{ (z_1^2\rb^2 + \mu_1^2)(z_2^2\rb^2 + \mu_2^2) } - \frac{z_1 \zb_2}{(z_1^2\rb^2 + \mu_1^2)(\zb_2^2\rb^2 + \mu_2^2)} \right. \nonumber
\\                                                                                                           
&&\hspace{-2.9cm}-\left. \frac{\zb_1 z_2}{(\zb_1^2\rb^2 + \mu_1^2)(z_2^2\rb^2 + \mu_2^2)} + \frac{\zb_1 \zb_2}{(\zb_1^2\rb^2 + \mu_1^2)(\zb_2^2\rb^2 + \mu_2^2)} \right] r^2 I_2 \nonumber
\\
&&\hspace{-2.90cm}+\left(\left[\frac{z_1}{z_1^2\rb^2+\mu_1^2}-\frac{\zb_1}{\zb_1^2\rb^2+\mu_1^2}\right]\left[I_{2\mu_2}(\zb_2)-I_{2\mu_2}(z_2)+r^2(\zb_2-z_2)I_{3\mu_2}(z_2)\right] + (1\lr2) \right) \nonumber
\\
&&\hspace{-2.9cm}+ r^2 \left[\frac{1}{2} + 2 z_1 z_2 -(z_1 + z_2)\right] \left[ I_{4\mu_1\mu_2}(z_1,\,z_2) + I_{4\mu_1\mu_2}(z_1,\,\zb_2) \right] \nonumber
\\
&&\hspace{-2.9cm}+ \left[- \frac{1}{2} + (z_1 + z_2)\right] I_{3\mu_1\mu_2}(z_1,\,z_2) + \left[ \frac{3}{2} - (z_1 + z_2)\right] I_{3\mu_1\mu_2}(\zb_1,\,\zb_2) \nonumber
\\
&&\hspace{-2.9cm}
+ \left[z_1 - z_2 -\frac{1}{2} \right] I_{3\mu_1\mu_2}(\zb_1,\,z_2) + \left[ z_2 - z_1 -\frac{1}{2} \right]  I_{3\mu_1\mu_2}(z_1,\,\zb_2)] \nonumber
\\
&&\hspace{-2.9cm}+ \frac{1}{r^2} [ J^c_{\mu_1 \mu_2}(z_1,\,\zb_2) + J^c_{\mu_1 \mu_2}(\zb_1,\,z_2) - J^c_{\mu_1 \mu_2}(\zb_1,\,\zb_2) - J^c_{\mu_1 \mu_2}(z_1,\,z_2) ]\,,
\eea
\bea
\label{f_old}
f (\rb;Q_1,Q_2;z_1,z_2) &=&\left[\frac{1}{z_1^2\rb^2+\mu_1^2}+\frac{1}{\zb_1^2\rb^2+\mu_1^2}\right]\left[\frac{z_2}{z_2^2\rb^2+\mu_2^2}-\frac{\zb_2}{\zb_2^2\rb^2+\mu_2^2}\right]  I_2 \nonumber
\\
&&\hspace{-3cm}+\, 2 \left[\frac{\zb_2}{\zb_2^2\rb^2+\mu_2^2}-\frac{z_2}{z_2^2\rb^2+\mu_2^2}\right] I_{3\mu_1}(z_1)  +(\zb_2-z_2)\left[\frac{1}{z_1^2\rb^2+\mu_1^2}+\frac{1}{\zb_1^2\rb^2+\mu_2^2}\right] I_{3\mu_2}(z_2 ) \nonumber
\\
&&\hspace{-4.2cm}+(z_2-\zb_2)\left[I_{4\mu_1\mu_2}(z_1,z_2 )\!+\!
I_{4\mu_1\mu_2}(\zb_1,z_2 )\right]\! + \! \frac{1}{r^2}\left[\frac{1}{z_1^2\rb^2+\mu_1^2}+\frac{1}{\zb_1^2\rb^2+\mu_1^2}\right][I_{2\mu_2}(\zb_2)\!-\! I_{2\mu_2}(z_2)] \nonumber
\\
&&\hspace{-3cm}+ \frac{1}{r^2}  [ I_{3\mu_1\mu_2}(\zb_1,\,z_2) -I_{3\mu_1\mu_2}(z_1,\,\zb_2)+I_{3\mu_1\mu_2}(z_1,\,z_2) - I_{3\mu_1\mu_2}(\zb_1,\,\zb_2)]\,,
\eqa
part of which are divergent integrals
\beqa
\label{I2}
I_2&=&\int \frac{d^d \kb}{\kb^2 (\kb - \rb)^2}\,, \\
\label{I2m}
I_{2m}(a)&=&\int \frac{d^d \kb}{\kb^2 ((\kb- a \rb)^2 + m^2)}\,, \\
\label{I3m}
I_{3m}(a)&=&\int \frac{d^d \kb}{\kb^2 (\kb - \rb)^2((\kb- a \rb)^2 + m^2)}\,, \\
\label{I3mm}
I_{3m_am_b}(a,\,b)&=&\int \frac{d^d \kb}{\kb^2((\kb-a \rb)^2 + m_a^2)((\kb-b \rb)^2 + m_b^2)}\,, \\
\label{I4mm}
I_{4m_am_b}(a,\,b)&=&\int \frac{d^d \kb}{\kb^2 (\kb - \rb)^2((\kb- a \rb)^2 + m_a^2)((\kb-b \rb)^2 + m_b^2)}\,,
\eqa
and some of them are finite 
\beqa
\label{J2mm}
J_{2 m_a m_b}(a,\,b)&=&\int \frac{d^d \kb}{((\kb- a \,\rb)^2 + m_a^2)((\kb- b \, \rb)^2 + m_b^2)}\,, \\
\label{Jc}
J^{c}_{m_a m_b}(a,\,b)&=&\int \frac{d^d \kb \ (\kb \cdot \rb) }{\kb^2 ((\kb- a \,\rb)^2 + m_a^2)((\kb-b \, \rb)^2 + m_b^2)}\,, \\
\label{K}
K_{m_a m_b}(a,\,b) &=&\int \frac{d^d \kb}{\kb^2}\left[ \frac{1}{(b^2 \rb^2  + m_b^2)[ (\kb-a \rb)^2 +m_a^2]}+ \frac{1}{(a^2 \rb^2  + m_a^2)[ (\kb-b \rb)^2 +m_b^2]} \right. \nonumber
\\
&&\left. -\frac{2 r^2}{[\kb^2+(\rb-\kb)^2](a^2\rb^2  + m_a^2)(b^2 \rb^2  + m_b^2)}\right] \,,\\
\label{L}
L_{m_a m_b}(a,\,b) &=&\int \frac{d^d \kb}{\kb^2}\left[  \frac{1}{(a^2\rb^2  + m_a^2)[ (\kb-b\rb)^2 +m_b^2]} \right. \\
&& \hspace{1cm} \left. -\frac{ r^2}{[\kb^2+(\rb-\kb)^2](a^2\rb^2 + m_a^2)(b^2\rb^2 + m_b^2)}\right] \,,\nonumber \\
\label{N}
N_{m_a m_b}(a,\,b) &=&\int \frac{d^d \kb}{\kb^2}\left[  \frac{1}{(a^2\rb^2  + m_a^2)[ (\kb-b\rb)^2 +m_b^2]}- \frac{1}{(b^2\rb^2 + m_b^2)[ (\kb-a \rb)^2 +m_a^2]} \right]\,,\nonumber \\
&& \\
\label{defJ3m}
J_{3 \mu}(a)&=& \int \frac{d^2 \kb}{\kb^2 (\kb - \rb)^2} \left
[ \frac{1}{(\kb - \rb a)^2 + \mu^2} -\frac{1}{a^2 \rb^2 + \mu^2}+ (a \lr
\aab) \right] \,, \\
\label{defJ4mm}
J_{4 \mu_1 \mu_2}(z_1,\, z_2)&=& \int \frac{d^2 \kb}{\kb^2
(\kb - \rb)^2} \\
&& \hspace{-1cm}\times \left[ \frac{1}{((\kb - \rb z_1)^2 + \mu_1^2)((\kb - \rb z_2)^2 + \mu_2^2)}
-\frac{1}{(z_1^2 \rb^2 + \mu_1^2)(z_2^2 \rb^2 +\mu_2^2)}+ (z \lr \zb) \right] \,. \nonumber
\eea

Although divergent integrals (\ref{I2}-\ref{I4mm}) enter formulas (\ref{a_old},\ref{b_old} and \ref{f_old}), 
the coefficients $a (\rb;Q_1,Q_2;z_1,z_2)$, $b (\rb;Q_1,Q_2;z_1,z_2)$ and $f (\rb;Q_1,Q_2;z_1,z_2)$ are finite expressions. This requires
a  tracing of the cancellation of divergencies within each of these 
coefficients which in principle can be done within a dimensional 
regularization scheme. Unfortunately this procedure is not always 
practically feasible, especially for the integrals $I_{3 m }$, $I_{3 m_1 m_2} $   or $  I_{4 m_1 m_2} $,
which apart of massless propagators contain also massive propagators.
These integrals can be rewritten in terms of simpler, although 
divergent integrals $I_2 $ and $ I_{2 m}$ as
\bea
\label{I3mmbis}
I_{3m_1m_2}(z_1,\,z_2)&=&-\frac{1}{2}\left(\frac{1}{z_1^2\rb^2+m_1^2}+\frac{1}{z_2^2\rb^2+m_2^2}\right) J_{2m_1 m_2}(z_1,\,z_2)
\\
&&\hspace{-4.35cm}+\!\left(\!\frac{z_1}{z_1^2\rb^2+m_1^2}\!+\!\frac{z_2}{z_2^2\rb^2+m_2^2}\right) \!J^{c}_{m_1 m_2}(z_1,\,z_2) \!+\!\frac{1}{2(z_2^2\rb^2+m_2^2)} I_{2 m_1}(z_1) \!+\! \frac{1}{2(z_1^2\rb^2+m_1^2)} I_{2 m_2}(z_2) \nonumber
 \,,
\\
\label{I4mmbis}
I_{4m_1m_2}(z_1,\,z_2)&=&\frac{1}{2}\left(\frac{1}{(z_1^2\rb^2+m_1^2)(z_2^2\rb^2+m_2^2)}+\frac{1}{(\zb_1^2\rb^2+m_1^2)(\zb_2^2\rb^2+m_2^2)}\right)  I_2  \nonumber \\
&&+\frac{1}{2} J_{4 m_1 m_2}(z_1,\, z_2) \,, \\
\label{I3mbis}
I_{3_m}(z)&=&\frac{1}{2}\left(\frac{1}{z^2 \rb^2+m^2}+\frac{1}{\zb^2\rb^2+m^2} \right) I_2 +\frac{1}{2} J_{3 m}(z) \,.
\eea
\\
Divergent  integrals like $I_2 $ and $ I_{2 m}$ or finite ones like $ J^{c}_{m_1 m_2}$ and $J_{2m_1 m_2}$ are directly computed by standard Feynman techniques.\\

The finite integrals $J_{4 m_1 m_2}$ and $ J_{3 m}$ are calculated by the use of 
the change of variables 
defined by the inverse conformal transformation applied to the momenta
 and other dimensional parameters of integrals, see [5] for details.
Such a change of variables applied to the two-dimensional, UV and 
IR finite integrals results in a reduction of number of massless 
propagators, a well known property from the studies of conformal field 
theories in coordinate space \cite{vassiliev}.


As a result, we get the following expressions for the coeffcients $a (\rb;Q_1,Q_2;z_1,z_2)$, $b (\rb;Q_1,Q_2;z_1,z_2)$ and $f (\rb;Q_1,Q_2;z_1,z_2)$ which involve only explicitely convergent integrals
\bea
\label{a}
 a (\rb,z_1,\,z_2) &=&  \\
&& \hspace{-2cm} \left( \frac{1}{2}\left[\frac{z_1}{ (z_1^2\rb^2 + \mu_1^2) } + \frac{z_2}{ (z_2^2\rb^2 + \mu_2^2) } +\frac{1}{r^2}\right] J^{c}_{\mu_1 \mu_2}(z_1,\,z_2)  +\frac{1}{4} K(z_1,\,z_2) \right.
\nonumber \\
&&\hspace{-2cm} \left. + \, (z_1 \lr \zb_1,\, z_2 \lr \zb_2 )- (z_1 \lr \zb_1) - (z_2 \lr \zb_2) \right) \nonumber \\
&&\hspace{-2cm}+\frac{r^2}{4} \left[J_{4 \mu_1 \mu_2}(\zb_1,\,z_2)-J_{4 \mu_1 \mu_2}(z_1,\,z_2) \right]+ \,
 \frac{1}{4}\, \left[ \frac{1}{ (z_1^2\rb^2 + \mu_1^2) } + \frac{1}{ (z_2^2\rb^2 + \mu_2^2) } +\frac{1}{ (\zb_1^2\rb^2 + \mu_1^2) }\right. \nonumber \\
&&\hspace{-2cm} \left. + \frac{1}{ (\zb_2^2\rb^2 + \mu_2^2) } \right] [J_{2\mu_1 \mu_2}(\zb_1,\,z_2)- J_{2\mu_1 \mu_2}(z_1,\,z_2)] \,, \nonumber \\ \\
\label{b}\nonumber
b(\rb,z_1,\,z_2)&=&  \\ &&\hspace{-2.7cm}\frac{r^2}{2}\left[\frac{z_1}{z_1^2\rb^2+\mu_1^2}-\frac{\zb_1}{\zb_1^2\rb^2+\mu_1^2}\right]\left[\zb_2-z_2 \right] J_{3 \mu_2}(z_2) + (1\lr2) \nonumber \\
&&\hspace{-2.8cm}+ r^2 \left[\frac{1}{2} + 2 z_1 z_2 -\!(z_1 + z_2)\right] \!\left[ J_{4 \mu_1 \mu_2}(z_1,\,z_2) + J_{4 \mu_1 \mu_2}(z_1,\,\zb_2) \right] \!+\! \left[- \frac{1}{2} + z_1 + z_2\right]\! \left[\!\left(\!\frac{z_1}{z_1^2\rb^2+\mu_1^2} \right.\right.\nonumber
\\
&&\hspace{-2.8cm} \left. \left. +\frac{z_2}{z_2^2\rb^2+\mu_2^2}\right) \!J^{c}_{\mu_1 \mu_2}(z_1,\,z_2) \!-\!\frac{1}{2}\left(\frac{1}{z_1^2\rb^2+\mu_1^2}+\frac{1}{z_2^2\rb^2+\mu_2^2}\right)\!J_{2\mu_1 \mu_2}(z_1,\,z_2)\right] \!+\!\left [ \frac{3}{2} - (z_1 + z_2)\right] \nonumber
\\
&&\hspace{-2.8cm} \times \, \left[\left(\frac{\zb_1}{\zb_1^2\rb^2+\mu_1^2}+\frac{\zb_2}{\zb_2^2\rb^2+\mu_2^2}\right) J^{c}_{\mu_1 \mu_2}(\zb_1,\,\zb_2) -\frac{1}{2}\left(\frac{1}{\zb_1^2\rb^2+\mu_1^2}+\frac{1}{\zb_2^2\rb^2+\mu_2^2}\right)J_{2\mu_1 \mu_2}(\zb_1,\,\zb_2)\right] \nonumber
\\
&&\hspace{-2.8cm}+\left [z_1 - z_2 -\frac{1}{2}\right] \left[\left(\frac{\zb_1}{\zb_1^2\rb^2+\mu_1^2}+\frac{z_2}{z_2^2\rb^2+\mu_2^2}\right) J^{c}_{\mu_1 \mu_2}(\zb_1,\,z_2) -\frac{1}{2}\left(\frac{1}{\zb_1^2\rb^2+\mu_1^2}+\frac{1}{z_2^2\rb^2+\mu_2^2}\right) \right. \nonumber \\ 
&&\hspace{-2.8cm} \left. \times \, J_{2\mu_1\mu_2}(\zb_1,\,z_2)\right] 
+ \left[z_2 - z_1 -\frac{1}{2}\right] \left[\left(\frac{z_1}{z_1^2\rb^2+\mu_1^2}+\frac{\zb_2}{\zb_2^2\rb^2+\mu_2^2}\right) J^{c}_{\mu_1 \mu_2}(z_1,\,\zb_2) -\frac{1}{2}\left(\frac{1}{z_1^2\rb^2+\mu_1^2} \right. \right.  \nonumber \\
&&\hspace{-2.8cm} \left. \left. +\frac{1}{\zb_2^2\rb^2+\mu_2^2}\right)\!J_{2\mu_1\mu_2}(z_1,\,\zb_2)\right] \!+ \! \frac{1}{r^2} \left [ J^{c}_{\mu_1 \mu_2}(z_1,\,\zb_2) \!+\! J^{c}_{\mu_1 \mu_2}(\zb_1,\,z_2) \!-\! J^{c}_{\mu_1 \mu_2}(\zb_1,\,\zb_2) \right. \nonumber \\
&&\hspace{-2.8cm} \left. - J^{c}_{\mu_1 \mu_2}(z_1,\,z_2) \right] + L_{\mu_1 \mu_2}(\zb_1,\,z_2) +L_{\mu_1 \mu_2}(\zb_2,\,z_1) -\frac{1}{4}[K_{\mu_1 \mu_2}(z_1,\,z_2)+K(\zb_1,\,\zb_2)  \nonumber \\
&&\hspace{-2.8cm}  +K_{\mu_1 \mu_2}(z_1,\,\zb_2)+K_{\mu_1 \mu_2}(\zb_1,\,z_2)] + \frac{z_2-z_1}{2}[N_{\mu_1 \mu_2}(z_1,\,z_2)+N_{\mu_1 \mu_2}(\zb_2,\,\zb_1)] \nonumber \\
&&\hspace{-2.8cm} +  \frac{z_1+z_2}{2}[N_{\mu_1 \mu_2}(z_1,\,\zb_2)+N_{\mu_1 \mu_2}(z_2,\,\zb_1)]
\,, \nonumber \\
\label{f}
f(\rb,z_1,\,z_2)&=&\left[\frac{\zb_2}{\zb_2^2\rb^2+\mu_2^2}-\frac{z_2}{z_2^2\rb^2+\mu_2^2}\right] J_{3 \mu_1}(z_1) 
\\
&&\hspace{-1cm}+\frac{z_2-\zb_2}{2}\left[ J_{4 \mu_1 \mu_2}(z_1,\,z_2) + J_{4 \mu_1\mu_2}(\zb_1,\,z_2) -(\frac{1}{z_1^2\rb^2+\mu_1^2}+\frac{1}{\zb_1^2\rb^2+\mu_1^2}) J_{3\mu_2}(z_2) \right] \nonumber
\\
&&\hspace{-1cm}+\left(\frac{1}{r^2}\left[-\frac{1}{2}(\frac{1}{z_1^2\rb^2+\mu_1^2}+ \frac{1}{z_2^2\rb^2+\mu_2^2})J_{2\mu_1 \mu_2}(z_1,\,z_2) +(\frac{z_1}{z_1^2\rb^2+\mu_1^2} \right. \right.\nonumber
\\
&&\hspace{-1cm}\left. \left. +\frac{z_2}{z_2^2\rb^2+\mu_2^2}) J^{c}_{\mu_1 \mu_2}(z_1,\,z_2)\right] -(z_1\lr \zb_1,\,z_2\lr \zb_2) +(z_1\lr \zb_1)-(z_2\lr \zb_2)  \right) \nonumber
\\
&&\hspace{-1cm}+\frac{1}{2 r^2} \left[ N(\zb_1,\,\zb_2)-N(z_1,\,z_2)+N(z_1,\,\zb_2)-N(\zb_1,\,z_2) \right]
 \,.\nonumber
\eea

The separation of divergent and finite parts in all integrals involved
was done within the dimensional regularization with  $d = 2 +2 \epsilon$.
Thus for the simplest integral $I_{2m}$ the result reads
\beq
\label{resultI2m}
I_{2m}(a)=\frac{\pi}{\eps(a^2\rb^2 +m^2)} (1 + \eps (\ln \pi - \Psi(1) - \ln
m^2+2 \ln(a^2\rb^2+m^2)))\,.
\eq
or for its massless case
\beq
\label{resultI2}
I_{2}= \frac{2 \pi}{\rb^2 \eps}(1 + \eps (\ln (\pi \rb^2)-\Psi(1)))\;.
\eq
We will use the finite part of $I_{2}$ and $I_{2m}$ to compute the integrals in which they are involved. For example we can write the finite integrals $K_{m_a m_b}$ , $L_{m_a m_b}$ and $M_{m_a m_b}$ in term of the finite part of these integrals since the divergent parts ultimately cancel when we express them like 
\bea
\label{tildeK}
K_{m_1 m_2}(z_1,\,z_2) &=&\frac{1}{z_2^2\rb^2+m_2^2} I_{2 m_1}(z_1) +\frac{1}{z_1^2\rb^2+m_1^2}I_{2m_2}(z_2) \nonumber
\\
 &&-\frac{r^2}{(z_1^2\rb^2+m_1^2)(z_2^2\rb^2+m_2^2)}I_2
 \,, \\
\label{tildeL}
L_{m_1 m_2}(z_1,\,z_2) &=& \frac{1}{z_1^2\rb^2+m_1^2}I_{2m_2}(z_2)-\frac{r^2}{2(z_1^2\rb^2+m_1^2)(z_2^2\rb^2+m_2^2)}I_2
 \,, \\
\label{tildeN}
N_{m_1 m_2}(z_1,\,z_2)&=&\frac{1}{z_1^2\rb^2+m_1^2}I_{2m_2}(z_2)-\frac{1}{z_2^2\rb^2+m_2^2} I_{2 m_1}(z_1)
 \,.
\eea
The integrals which involve only two massive propagators are finite
and are calculated in a standard way using Feynman parameters. We obtain

\bea
\label{tildeI2mm}
J_{2m_a m_b}&=&\int \frac{d^d \kb}{((\kb-\rb a)^2+m_a ^2)
((\kb-\rb b)^2+m_b^2)}\nonumber \\
&=&\frac{\pi}{\sqrt{\lambda}}  \ln \frac{r^2(a-b)^2 +m_a^2
+m_b^2 +\sqrt{\lambda}}{r^2(a-b)^2 +m_a^2
+m_b^2 -\sqrt{\lambda}}\,.
\eea
where we introduce the notation
\beq
\label{deffonctionlambda}
\lambda(x,y,z)=x^2+y^2+z^2-2 x y -2 x z -2 y z\,,
\eq
which  enables us to define, for the purpose of our computation,
\beq
\label{deflambda}
\lambda=\lambda(-r^2 (a-b)^2,\alpha^2,\beta^2)=(\alpha^2 -\beta^2)^2 +2(\alpha^2+\beta^2)r^2 (a-b)^2 + r^4
(a-b)^4\,.
\eq
For $J_{m_a m_b}^c$ we get
\bea
\label{tildeJcmm}
\hspace{-1cm}&&J_{m_a m_b}^c=\frac{2 \pi  r^2(a-b)}{\sqrt{\lambda}}\left[\frac{1}{\sqrt{\lambda}+m_a^2 -m_b^2+r^2(a^2-b^2)}  \ln \frac{a (\sqrt{\lambda}+r^2(a-b)^2 +m_a^2-m_b^2)}{b (\sqrt{\lambda}-r^2(a-b)^2 +m_a^2-m_b^2)} \right.\nonumber
\\
&&\left. +\frac{1}{\sqrt{\lambda}+m_b^2 -m_a^2+r^2(b^2-a^2)}  \ln \frac{a (\sqrt{\lambda}-r^2(a-b)^2 -m_a^2+m_b^2)}{b (\sqrt{\lambda}+r^2(a-b)^2 -m_a^2+m_b^2)}\right] \nonumber
\\
&&- \frac{\pi}{a(b^2r^2+m_b^2)-b(a^2r^2+m_a^2)}  \ln \frac{a(b^2r^2+m_b^2)}{b(a^2r^2+m_a^2)}
\,.
\eea
The $ \ln\frac{a}{b}$ terms which apparently give divergencies when a or b goes to zero are spurious: indeed they cancel because of the symmetrical way the function $J_{m_a m_b}^c$ appears in $a(\rb;Q_1,Q_2;z_1,z_2)$, $b(\rb;Q_1,Q_2;z_1,z_2)$ or $f(\rb;Q_1,Q_2;z_1,\, z_2)$.

Let us now consider the $J_{3m}$ integral.
After performing a special conformal transformation, namely an inversion on momentum integration variables and 
other dimensional vectors and parameters, a translation and again an  
inversion, we arrive to integrals with smaller 
number of propagators which are calculated in the standard way. The final 
result reads
\bea
\label{resultJ3m}
&&\hspace{-.5cm}J_{3 m} = \frac{2 \pi}{r^2} \left\{ \left(\frac{1}{r^2 a^2 + m^2}-\frac{1}{r^2 \aab^2 + m^2}
\right) \ln \frac{r^2 a^2 +m^2}{r^2 \aab^2 + m^2} \right.\nonumber \\
&&\hspace{-.5cm}\left.+\left(\frac{1}{r^2 a^2
+ m^2}
+\frac{1}{r^2 \aab^2 + m^2}+ \frac{2}{r^2 a \aab - m^2}\right) \ln
\frac{(r^2 a^2 + m^2)(r^2 \aab^2 + m^2)}{m^2 r^2} \right\}\,.
\eea
With the same  but more tricky approach we can compute the $J_{4mm}$  integral after performing a special conformal transformation; we refer to \cite{psw} for the complete calculation and final expression of this integral (see (A.66) and (A.67) of \cite{psw}). Starting from this result, we now derive another expression in such way that some spurious divergent terms (appearing when $r^2=Q_i^2\,,$ which corresponds to $a=0$ or $b=0$) explicitely cancel. Thus it allows us to use it in our numerical integration code (cf. sections \ref{cross} and \ref{crossee}). This makes the explicit expression (\ref{resJ3finie}) different from the one given in (A.67) of \cite{psw}.
The resulting expression for $J_{4 \mu_1 \mu_2}$ is
\bea
\label{resI}
J_{4 \mu_1 \mu_2}= \hspace{-.2cm}\\
&&\hspace{-2.1cm} \frac{1}{r^2 (r^2 \zb_1^2 + \mu_1^2) (r^2 \zb_2^2 + \mu_2^2)}J^{finite}_{3
  \alpha \beta}\!\left(\frac{-z_1 \zb_1 r^2 +\mu_1^2}{\zb_1^2 r^2+\mu_1^2},
\frac{-z_2 \zb_2 r^2 +\mu_2^2}{\zb_2^2 r^2+\mu_2^2}, \frac{r^2 \mu_1}{r^2
  \zb_1^2 + \mu_1^2}, \frac{r^2 \mu_2}{r^2
  \zb_2^2 + \mu_2^2}, r \right) \nonumber \\
&&\hspace{-2.1cm}+ \frac{1}{r^2 (r^2 z_1^2 + \mu_1^2) (r^2 z_2^2 + \mu_2^2)}J^{finite}_{3
  \alpha \beta}\!\left(\frac{-z_1 \zb_1 r^2 +\mu_1^2}{z_1^2 r^2+\mu_1^2},
\frac{-z_2 \zb_2 r^2 +\mu_2^2}{z_2^2 r^2+\mu_2^2}, \frac{r^2 \mu_1}{r^2
  z_1^2 + \mu_1^2}, \frac{r^2 \mu_2}{r^2
  z_2^2 + \mu_2^2}, r \right) \nonumber
\eea
with
\bea
\label{resJ3finie}
&&\hspace{-.5cm}J^{finite}_{3\alpha \beta}(a,\,b,\, \alpha,\, \beta,\, r)  =\hspace{-.5cm}\nonumber \\
&&\hspace{-.2cm}\frac{\pi}{4} \left\{ \left( \frac{8}{a - b}+\frac{2\,\left( {\alpha }^2 - {\beta }^2 \right) }{{\left( a - b \right) }^2\,r^2} \right) \ln \frac{{\alpha }^2}{{\beta }^2} +\frac{4r^2\left(-4 + r^2\left( \frac{a}{a^2\,r^2 + {\alpha }^2} + \frac{b}{b^2\,r^2 + {\beta }^2} \right)  \right) \,
    \ln \frac{b^2\,r^2 + {\beta }^2}{a^2\,r^2 + {\alpha }^2}}{a\,\left( a - b \right) \,b\,r^2 + b\,{\alpha }^2 - 
    a\,{\beta }^2}\right. \nonumber \\
&&\hspace{-.7cm} \left.  -2\,\ln \frac{{\alpha }^2\,{\beta }^2}{r^4} + \frac{2\,r^4}{\left( a^2\,r^2 + {\alpha }^2 \right) \,\left( b^2\,r^2 + {\beta }^2 \right) } \left( \ln \frac{{\alpha }^2\,{\beta }^2}{r^4} - 2\,\ln \frac{{\alpha }^2\,{\beta }^2}
     {\left( a^2\,r^2 + {\alpha }^2 \right) \,\left( b^2\,r^2 + {\beta }^2 \right) }\right) \right. \nonumber\\ 
&&\hspace{-.7cm} \left. -\frac{8}{\left( a - b \right) \,\left( a\,\left( a - b \right) \,b\,r^2 + b\,{\alpha }^2 - a\,{\beta }^2 \right) \, \left( - a^2\,r^2  +b^2\,r^2 -{\alpha }^2 + {\beta }^2 +{\sqrt{\lambda }}\right)}   \right. \nonumber\\ 
&&\hspace{-.7cm}  \times \, \left. \left[  \left( - {\left( a - b \right) }^2 \, r^2   - {\alpha }^2 + {\beta }^2 + {\sqrt{\lambda }} \right) \,
  \left(\frac{-a\left( b^2\,r^2 + {\beta }^2 \right) }{b}\,\ln \left(1 + \frac{b^2\,r^2}{{\beta }^2}\right)  
\right.\right.\right.  \\
&&\left.\left.\left.
+ 
 \left( a^2\,r^2 + {\alpha }^2 \right) \,\ln \frac{a^2\,r^2 + {\alpha }^2}{{\beta }^2} \right)  + \left( {\left( a - b \right) }^2\,r^2 - {\alpha }^2 + {\beta }^2 + {\sqrt{\lambda }} \right) \right. \right. \nonumber \\
&& \left. \left. \times \, \left(  \frac{-b\left( a^2\,r^2 + {\alpha }^2 \right) }{a}
\ln \left(1 + \frac{a^2\,r^2}{{\alpha }^2}\right)  + 
 \left( b^2\,r^2 + {\beta }^2 \right) \,\ln \frac{b^2\,r^2 + {\beta }^2}{{\alpha }^2} \right)    \right]  \right. \nonumber \\
 &&\hspace{-.7cm} \left. +\left( \frac{{\left( a^2\,r^2 - b^2\,r^2 + {\alpha }^2 - {\beta }^2 + {\sqrt{\lambda }} \right) }^2}
 {{\left( a - b \right) }^2\,r^2\,{\sqrt{\lambda }}} +  \frac{8\,\left( a - b \right) \,r^4\,\left( -4 +   r^2\,\left( \frac{a}{a^2\,r^2 + {\alpha }^2} + \frac{b}{b^2\,r^2 + {\beta }^2} \right)  \right) }{ \left( \left( a^2 - b^2 \right) \,r^2 + {\alpha }^2 - {\beta }^2 + {\sqrt{\lambda }} \right) \,{\sqrt{\lambda }}} \right. \right. \nonumber \\
&&\hspace{-.7cm} \left. \left. +\frac{8\,{\sqrt{\lambda }}}
  {\left( a - b \right) \,\left( a\,\left( a - b \right) \,b\,r^2 + b\,{\alpha }^2 - a\,{\beta }^2 \right) } \right) \ln \frac{{\left( a - b \right) }^2\,r^2 + {\alpha }^2 - {\beta }^2 + {\sqrt{\lambda }}}
   {- {\left( a - b \right) }^2\,r^2   + {\alpha }^2 - {\beta }^2 + {\sqrt{\lambda }}}  \right.\nonumber \\
&&\hspace{-.7cm}+ \!\left. \left(\! \! \frac{-{\left( - a^2\,r^2 + b^2\,r^2 - {\alpha }^2 + {\beta }^2 + {\sqrt{\lambda }} \right) }^2}
       {{\left( a - b \right) }^2\,r^2\,{\sqrt{\lambda }}}  + 
    \frac{8\,\left( a - b \right) \,r^4\,\left( -4 + 
         r^2\,\left( \frac{a}{a^2\,r^2 + {\alpha }^2} + \frac{b}{b^2\,r^2 + {\beta }^2} \right)  \right) }{
       \left( \left( -a^2 + b^2 \right) \,r^2 - {\alpha }^2 + {\beta }^2 + {\sqrt{\lambda }} \right) \,{\sqrt{\lambda }}}
    \!\! \right)  \right. \!\!\nonumber \\
&&\hspace{-.7cm}\times  \left. \ln \frac{- {\left( a - b \right) }^2\,r^2 - {\alpha }^2 + {\beta }^2 + {\sqrt{\lambda }}}{{\left( a - b \right) }^2\,r^2 - {\alpha }^2 + {\beta }^2 + {\sqrt{\lambda }}}  +\left[ \frac{16\,r^2}{{\sqrt{\lambda }}} + \frac{2\,r^2\,\left( 4 + 
       r^2\,\left( -\left( \frac{1}{a^2\,r^2 + {\alpha }^2} \right)  - \frac{1}{b^2\,r^2 + {\beta }^2} \right)  \right) }{
      {\sqrt{\lambda }}}\right.  \right. \nonumber \\
&&\hspace{-.7cm} \left. \left. \, - \, 8 \frac{ \left( a^2 - b^2 \right) \,r^2 + {\alpha }^2 - {\beta }^2  }{\left( a - b \right) \,{\sqrt{\lambda }}} - \frac{2\,{\sqrt{\lambda }}}{{\left( a - b \right) }^2\,r^2} \right] \ln \frac{{\left( a - b \right) }^2\,r^2 + {\alpha }^2 + {\beta }^2 + {\sqrt{\lambda }}}
   {{\left( a - b \right) }^2\,r^2 + {\alpha }^2 + {\beta }^2 - {\sqrt{\lambda }}}  \right\} \, , \hspace{-.5cm}\nonumber
\eea
 where  appropriate additional $\ln r^2$ terms have been introduced in order to write the final result
as made of logarithms of dimensionless quantities. The difference between these two expressions
only involves terms proportional to
$\ln r^2,$ in accordance to the dimensional regularization, since $J_{3 \alpha \beta}$ is a divergent integral. However, at the level of the
  final result for $J_{4 \mu_1 \mu_2},$ which is both UV and IR finite, these additional terms of course cancel each other.



\end{document}